\documentclass[aip,jcp,reprint,amsmath,amssymb]{revtex4-2}

\usepackage{amsmath,amssymb}
\usepackage{braket}
\usepackage{booktabs}
\usepackage{graphicx}
\usepackage{xr-hyper}
\IfFileExists{si.aux}{\externaldocument{si}}{\AtBeginDocument{
\newlabel{si:geometries}{{S1}{}{}{}{}}
\newlabel{si:h2o2}{{S2}{}{}{}{}}
\newlabel{tab:si_h2o2_mp2}{{S1}{}{}{}{}}
\newlabel{tab:si_h2o2_cisd}{{S2}{}{}{}{}}
\newlabel{si:smo}{{S3}{}{}{}{}}
\newlabel{fig:si_smo_pvdz}{{S1}{}{}{}{}}
\newlabel{fig:si_smo_apvdz}{{S2}{}{}{}{}}
\newlabel{tab:si_smo_mp2}{{S3}{}{}{}{}}
\newlabel{tab:si_smo_cisd}{{S4}{}{}{}{}}
\newlabel{si:cby}{{S4}{}{}{}{}}
\newlabel{fig:si_cby_pvdz}{{S3}{}{}{}{}}
\newlabel{fig:si_cby_apvdz}{{S4}{}{}{}{}}
\newlabel{tab:si_cby_mp2}{{S5}{}{}{}{}}
\newlabel{tab:si_cby_cisd}{{S6}{}{}{}{}}
\newlabel{si:sumrule}{{S5}{}{}{}{}}
\newlabel{tab:si_sumrule}{{S7}{}{}{}{}}
}}
\usepackage{hyperref}
\usepackage{xcolor}
\usepackage{mhchem}
\usepackage{braket}

\def\crt#1{a_{#1}^{\dagger}}
\def\ann#1{a_{#1}^{\ }}
\def\ket#1{| #1 \rangle}
\def\bra#1{\langle #1 |}
\def\bm#1{\mbox{\boldmath $#1$}}
\def\int#1#2{\left\langle #1 | #2 \right\rangle}
\def\smo{(\textit{S})-methyloxirane}
\def\cby{(\textit{S})-3-chloro-1-butyne}
\def\runit{10$^{-44}$ esu$^{2}$ cm$^{2}$}

\newcommand{\Pvg}{\vec{P}^\text{VG}_i}
\newcommand{\Plg}{\vec{P}^\text{LG}_i}
\newcommand{\M}{\vec{M}_i}
\newcommand{\Rvg}{R^\text{VG}_i}
\newcommand{\Rlg}{R^\text{LG}_i}
\newcommand{\Rlgoi}{R^\text{LG(OI)}_i}
\newcommand{\Dvg}{D^\text{VG}_i}
\newcommand{\Dlg}{D^\text{LG}_i}
\newcommand{\Dmix}{D^\text{LG/VG}_i}
\newcommand{\Di}{\left[\bm{D}_i\right]^\text{LG/VG}}
\newcommand{\Ri}{\left[\bm{R}_i\right]^\text{LG}}

\begin{document}

\title{A Density-Matrix Formulation of Vibrational Circular Dichroism Spectra for M{\o}ller-Plesset and Configuration Interaction Methods}
\author{Aparna Krishnan}
    \affiliation{Department of Chemistry, Virginia Tech, Blacksburg, VA 24061, U.S.A.}
\author{Brendan M. Shumberger}
    \affiliation{Department of Chemistry, Virginia Tech, Blacksburg, VA 24061, U.S.A.}
\author{T. Daniel Crawford}
    \affiliation{Department of Chemistry, Virginia Tech, Blacksburg, VA 24061, U.S.A.}
    \email{crawdad@vt.edu}
\date{\today}

\begin{abstract}
We reformulate the overlap of wave-function derivatives that defines the atomic
axial tensors (AATs) and velocity-gauge atomic polar tensors (VG-APTs) used to
compute vibrational circular dichroism (VCD) rotatory strengths. In place of our
earlier determinant-level derivatives, the new approach expresses these tensors
as contractions involving molecular-orbital derivative integrals, perturbed and
unperturbed one- and two-electron density matrices, and dot products of the
perturbed correlated wave-function amplitudes. Following the elegant treatment of the
diagonal Born-Oppenheimer correction by Gauss and co-workers, the formulation
holds for any real, normalized CI-like wave function. The two-electron density
contribution vanishes exactly, because one perturbation is real (nuclear) and the
other imaginary (magnetic field or momentum); this removes multiple terms from our
earlier AAT and VG-APT expressions for second-order M\o{}ller-Plesset
perturbation theory (MP2) and configuration interaction singles and doubles
(CISD). We also present a simplified form of the origin-independent length-gauge
[LG(OI)] rotatory strength that avoids the singular-value decomposition of the
outer product of the length-gauge APT and VG-APT vectors for each mode, and that
clarifies how the LG(OI) and velocity-gauge rotatory strengths are related.
Applying the density-based AATs and VG-APTs to ($P$)-hydrogen peroxide, \smo, and
\cby, we find that discrepancies among the gauges arise chiefly from basis-set
incompleteness in the VG-APT. The charge-sum rule serves as a useful measure of
that incompleteness, independent of the correlation method.
\end{abstract}

\maketitle

\section{Introduction}
\label{sec:intro}

Vibrational circular dichroism (VCD), the differential absorption of left- and
right-circularly polarized infrared radiation by chiral molecules, is among
the most powerful spectroscopic tools for assigning the absolute configuration
of molecules.\cite{Stephens1985, Nafie1997, Freedman2003, Barron04,
Polavarapu2007, Crawford2006review} VCD band intensities of fundamental
($0\rightarrow 1$) transitions in the electronic ground state for a given
vibrational mode $i$ are given by the rotatory strength,
\begin{equation}
\label{eq:VCD-def}
R(0 \rightarrow 1)_i = \textrm{Im} \left[ \braket{0 | \vec{\mu} | 1_i} \cdot
\braket{1_i | \vec{m} | 0} \right]
\end{equation}
where $\vec{\mu}$ and $\vec{m}$ are the electric- and magnetic-dipole
operators. The electric-dipole vibrational transition moment is
obtained from the atomic polar tensor (APT), which is conventionally computed as the
derivative of the ground-state dipole moment expectation value in the
electrical harmonic approximation [the length-gauge APT (LG-APT)]. The magnetic-dipole transition moment
(the atomic axial tensor, or AAT), on the other hand, cannot be computed using
the same approach because the corresponding magnetic-dipole expectation value
is zero in the Born-Oppenheimer approximation, and overcoming this deficiency
is the central computational challenge in simulating VCD spectra.

Most common implementations of the AAT are based on Stephens's magnetic-field
perturbation theory (MFPT) approach,\cite{Stephens1985} which requires
computing the overlap of wave function derivatives with respect to nuclear
coordinates and with respect to an external magnetic field.  Following
Stephens's foundational work in the mid-1980s, VCD implementations were
developed at the Hartree-Fock (HF),\cite{Stephens1985annrev,Lowe1986,Amos1987}
multiconfigurational self-consistent field (MCSCF),\cite{Bak1993,Bak1994} and
density functional theory (DFT)\cite{Cheeseman1996} levels.  Yang and Rauk
computed second-order M{\o}ller-Plesset (MP2) AATs within Nafie and Freedman's
vibronic-coupling theory,\cite{Nafie1983} using a sum over configuration
interaction singles (CIS) excited states,\cite{Yang1994} but MFPT-based AATs
from dynamically correlated wave functions have only recently appeared.  In
2024, Shumberger and Crawford reported finite-difference implementations at
the MP2 and CI doubles (CID) levels of theory,\cite{Shumberger2024jctc}
followed in 2025 by an analytic-derivative MP2 implementation using a
second-quantized orbital-derivative approach.\cite{Shumberger2025jctc} Using
the same approach, in 2026 they reported the extension of this analytic
framework to CID and CI singles and doubles (CISD),\cite{Shumberger2026jpca}
including the use of non-canonical perturbed orbitals and frozen-core
orbitals.

The determinant-based approach of Refs.~\onlinecite{Shumberger2025jctc} and
\onlinecite{Shumberger2026jpca} expresses the AAT through explicit
contractions involving CI coefficients, their derivatives, and overlaps
between perturbed Slater determinants.  While correct and carefully validated
against finite-difference results, this formulation yields a large number of
terms (e.g., 25 for the CISD AAT), particularly in the orbital-orbital block,
where numerous distinct contractions arise from the coupling between nuclear
and magnetic orbital responses.

In the present work, we reformulate the AAT and velocity-gauge APT (VG-APT)
for any CI-like (\textit{i.e.}, a linear expansion of determinants) wave
function in terms of perturbed and unperturbed one- and two-electron density
matrices, motivated by a structural analogy with the diagonal Born-Oppenheimer
correction (DBOC) as formulated by Gauss \textit{et al.}\cite{Gauss2006} We
show that an analogous decomposition applies to the AAT and VG-APT, with the
key distinction that the perturbations are nuclear-coordinate and
magnetic-field/linear-momentum rather than both nuclear.  A consequence of
this reformulation is that the two-electron density matrix contribution to the
AAT and VG-APT vanishes identically (in agreement with the finding of Bak
\textit{et al.} for the AAT for MCSCF wave functions\cite{Bak1993,Bak1994}), providing both
conceptual simplification and a pathway toward an efficient method-agnostic
generalization.

\section{Theory}

In this work we focus on the AAT, which is essential to simulations of VCD
spectra, and the VG-APT, which is relevant to both IR absorption and VCD
spectra.  Using Stephens's general MFPT formalism, the electronic component of
the AAT (with the constants that relate it to the magnetic-dipole transition
moment given in Ref.~\onlinecite{Stephens1985}) may be expressed as an overlap
of ground-state wave function derivatives, \textit{viz.},
\begin{equation}
I^\lambda_{\alpha\beta} = \int{\partial_{R_{\lambda\alpha}}
\Psi}{\partial_{H_\beta} \Psi}_0,
\end{equation}
where $\alpha$ and $\beta$ denote Cartesian directions of the nuclear
coordinates ($\vec{R}$) and external magnetic field ($\vec{H}$), respectively,
while the superscript $\lambda$ indexes the nuclei.  
Similarly, the electronic
contribution to the VG-APT may be written as 
\begin{equation}
    [P_{\alpha\beta}^\lambda]^{\text{VG}}
    = - \frac{2 i e \hbar}{m} \int{\partial_{R_{\lambda\alpha}} \Psi}
    {\partial_{A_\beta} \Psi}_0,
\end{equation}
where the vector potential, $\vec{A}$, enters the Hamiltonian through the
spatially uniform potential operator $\vec{A} \cdot \vec{\pi}^e$, where
$\vec{\pi}^e$ is the total electronic linear momentum.  In both the AAT and
the VG-APT, the subscript $0$ indicates that derivatives are taken at a
reference (equilibrium) geometry and at zero field ($H_\beta = 0$ or $A_\beta
= 0$). Each is proportional to a Berry curvature of the electronic ground state, taken with respect to 
the nuclear coordinates and either the magnetic field (for the AAT) or the vector 
potential (for the VG-APT).\cite{pausch2025berry, resta2023berry}
These expressions may also be considered generalizations of the conventional
formulation of the DBOC, which
accounts for the effect of nuclear kinetic energy operating on the electronic
wave function within the same electronic state, \textit{i.e.},
\begin{equation}
        \Delta E_\textrm{DBOC} = \sum_{\lambda\alpha} \frac{\hbar^2}{2M_\lambda} 
        \int{\partial_{R_{\lambda\alpha}} \Psi}
        {\partial_{R_{\lambda\alpha}} \Psi},
\end{equation}
where $M_\lambda$ is the mass of the $\lambda$-th nucleus.\cite{Kutzelnigg97:adiabatic}
While the DBOC involves identical real perturbations in both the bra and ket,
the AAT and VG-APT require real and imaginary perturbations.

\subsection{Density-Based Formulation}

Given a normalized, real CI-type wave function (\textit{i.e.}, a linear expansion of Slater determinants)
\begin{equation}
\Psi = \sum_I c_I \Phi_I,
\end{equation}
our goal is to compute the overlap of wave function derivatives with respect to external parameters $x$ and $y$, which may be real or imaginary,
\begin{equation}
S_{xy} = \int{\partial_x \Psi}{\partial_y \Psi}.
\label{Eq:dx_Psi_dy_Psi}
\end{equation}
Following Gauss \textit{et al.},\cite{Gauss2006} we decompose the derivatives into components acting on CI coefficients
and components acting on the molecular orbitals (MOs), \textit{i.e.},
\begin{equation}
\partial_x = \partial_x^c + \partial_x^\phi.
\end{equation}
The latter may be expressed in second quantization as
\begin{equation}
\partial_x^\phi = \sum_{pq} \int{\phi_p}{\partial_x \phi_q} \crt{p}\ann{q},
\end{equation}
where the anti-Hermitian matrix is
\begin{equation}
\int{\phi_p}{\partial_x \phi_q} = U^x_{pq} + \tilde{S}^x_{pq} = -\int{\partial_x \phi_p}{\phi_q},
\end{equation}
and $U^x_{pq}$ is a first-order coupled-perturbed Hartree-Fock (CPHF) coefficient, and $\tilde{S}^x_{pq}$ is the
half-derivative atomic-orbital (AO) overlap transformed into the MO basis.  Note that
\begin{equation}
\left(\partial_x^\phi\right)^\dagger = \sum_{pq} \int{\partial_x \phi_p}{\phi_q} \crt{p}\ann{q}.
\end{equation}
Inserting these into the overlap of wave function derivatives, we obtain
\begin{align}
S_{xy} &= 
\int{\partial_x^c \Psi}{\partial_y^c \Psi} + \int{\partial_x^c \Psi}{\partial_y^\phi \Psi} \nonumber \\
& \qquad +
\int{\partial_x^\phi \Psi}{\partial_y^c \Psi} + \int{\partial_x^\phi \Psi}{\partial_y^\phi \Psi}.
\end{align}
The first term is written as
\begin{equation}
\int{\partial_x^c \Psi}{\partial_y^c \Psi} = \sum_{I} (c_I^x)^* c_I^y,
\end{equation}
where the superscripts indicate the derivatives of the $c_I$ with respect to $x$ or $y$. 
The second term becomes
\begin{equation}
\int{\partial_x^c \Psi}{\partial_y^\phi \Psi} = \sum_{IJ} (c_I^x)^* c_J \sum_{pq} \int{\phi_p}{\partial_y \phi_q} \bra{\Phi_I} \crt{p}\ann{q} \ket{\Phi_J},
\end{equation}
which may be written in terms of a half-perturbed one-electron density,
\begin{equation}
D^{(x,0)}_{pq} \equiv \sum_{IJ} (c_I^x)^* c_J \bra{\Phi_I} \crt{p}\ann{q} \ket{\Phi_J}
\end{equation}
to give
\begin{equation}
\int{\partial_x^c \Psi}{\partial_y^\phi \Psi} = \sum_{pq} D^{(x,0)}_{pq} \int{\phi_p}{\partial_y \phi_q}.
\end{equation}
Similarly, the third term becomes
\begin{equation}
\int{\partial_x^\phi \Psi}{\partial_y^c \Psi} = \sum_{pq} D^{(0,y)}_{pq} \int{\partial_x \phi_p}{\phi_q}
\end{equation}
where
\begin{equation}
D^{(0,y)}_{pq} \equiv \sum_{IJ} c_I^* c_J^y \bra{\Phi_I} \crt{p}\ann{q} \ket{\Phi_J}.
\end{equation}
Finally, the fourth term may be written as
\begin{align}
\int{\partial_x^\phi \Psi}{\partial_y^\phi \Psi} &= 
\sum_{IJ} c_I^* c_J \sum_{pq} \int{\partial_x \phi_p}{\partial_y \phi_q} \bra{\Phi_I} \crt{p}\ann{q} \ket{\Phi_J} \nonumber \\
& \hspace{-4em} + \sum_{IJ} c_I^* c_J \sum_{pqrs} \int{\partial_x \phi_p}{\phi_q} \bra{\Phi_I} \crt{p}\crt{r} \ann{s}\ann{q} \ket{\Phi_J} \int{\phi_r}{\partial_y \phi_s},
\end{align}
where the first term on the right-hand side arises from either (1) the fact
that the derivative operators can produce perturbations orthogonal to the
unperturbed MO space (\textit{i.e.}, $x$ and $y$ MO derivatives may be
non-orthogonal to each other) or more simply, (2) the anti-commutation
relations of the annihilation/creation operators in a complete MO basis.
Defining unperturbed one- and two-electron densities as
\begin{equation}
D^{(0,0)}_{pq} \equiv \sum_{IJ} (c_I)^* c_J \bra{\Phi_I} \crt{p}\ann{q} \ket{\Phi_J}
\end{equation}
and
\begin{equation}
        \Gamma^{(0,0)}_{prqs} \equiv \sum_{IJ} (c_I)^* c_J \bra{\Phi_I} \crt{p}\crt{r}\ann{s}\ann{q} \ket{\Phi_J},
\end{equation}
respectively, yields
\begin{align}
\int{\partial_x^\phi \Psi}{\partial_y^\phi \Psi} &= 
\sum_{pq} D^{(0,0)}_{pq} \int{\partial_x \phi_p}{\partial_y \phi_q} \nonumber \\ 
&+ \sum_{pqrs} \Gamma^{(0,0)}_{prqs} \int{\partial_x \phi_p}{\phi_q} \int{\phi_r}{\partial_y \phi_s}.
\end{align}
Thus, the complete expression for the overlap in Eq.~\eqref{Eq:dx_Psi_dy_Psi}
for a real, linear wave function is
\begin{align}
S_{xy} &= \sum_{I} (c_I^x)^* c_I^y + \sum_{pq} D^{(x,0)}_{pq} \int{\phi_p}{\partial_y \phi_q} \nonumber \\
& + \sum_{pq} D^{(0,y)}_{pq} \int{\partial_x \phi_p}{\phi_q} + \sum_{pq} D^{(0,0)}_{pq} \int{\partial_x \phi_p}{\partial_y \phi_q}
\nonumber \\
& + \sum_{pqrs} \Gamma^{(0,0)}_{prqs} \int{\partial_x \phi_p}{\phi_q} \int{\phi_r}{\partial_y \phi_s}
\label{Eq:final_overlap}
\end{align}

\subsection{Symmetry and the AAT/VG-APT}

While Eq.~\eqref{Eq:final_overlap} is general for the choice of perturbations
$x$ and $y$, the symmetries of its component terms play a key role in specific
applications.  For example, if $x$ is a real perturbation (\textit{e.g.}, a nuclear
coordinate), then the half-derivative MO overlap is antisymmetric, \textit{i.e.},
\begin{equation}
\int{\partial_x \phi_p}{\phi_q} = -\int{\partial_x \phi_q}{\phi_p},
\end{equation}
but if $x$ is imaginary (\textit{e.g.}, a magnetic field or linear momentum),
then the half-derivative overlap is symmetric,
\begin{equation}
\int{\partial_x \phi_p}{\phi_q} = \int{\partial_x \phi_q}{\phi_p}.
\end{equation}
In addition, for a real CI-like wave function, the two-electron density exhibits bra-ket symmetry, that is
\begin{equation}
\Gamma^{(0,0)}_{prqs} = \Gamma^{(0,0)}_{qspr}.
\end{equation}
(Note that this is not the case for a coupled cluster wave function due to the
conventional lack of Hermiticity of the similarity-transformed Hamiltonian, though the present formulation does not
directly apply to biorthonormal wave functions.)
In the case that $x$ is real and $y$ is imaginary (such as for the AAT or VG-APT), the final term of Eq.~\eqref{Eq:final_overlap} vanishes because
\begin{align}
\sum_{pqrs} \Gamma^{(0,0)}_{prqs} \int{\partial_x \phi_p}{\phi_q} \int{\phi_r}{\partial_y \phi_s} & \nonumber \\
& \hspace{-8em} = \sum_{pqrs} \Gamma^{(0,0)}_{qspr} \int{\partial_x \phi_q}{\phi_p} \int{\phi_s}{\partial_y \phi_r} \nonumber \\
& \hspace{-8em} = -\sum_{pqrs} \Gamma^{(0,0)}_{prqs} \int{\partial_x \phi_p}{\phi_q} \int{\phi_r}{\partial_y \phi_s} \nonumber \\
& \hspace{-8em} = 0.
\end{align}
Thus, in the case of the AAT and VG-APT, no explicit two-electron-density terms contribute, reducing the expression to 
\begin{align}
S_{xy} & = \sum_{I} (c_I^x)^* c_I^y + \sum_{pq} D^{(x,0)}_{pq} \int{\phi_p}{\partial_y \phi_q} \nonumber \\
& + \sum_{pq} D^{(0,y)}_{pq} \int{\partial_x \phi_p}{\phi_q} + \sum_{pq} D^{(0,0)}_{pq} \int{\partial_x \phi_p}{\partial_y \phi_q}
\label{Eq:real_imag_overlap}
\end{align}
in agreement with previous SCF,\cite{Amos88} DFT,\cite{Cheeseman1996} and
MCSCF\cite{Bak1993,Bak1994} implementations for the AAT.  Note that this reduction is
solely based on the parity of the relevant half-derivative overlap integrals,
not on the use of magnetic-field or nuclear-velocity-dependent basis
functions.  Bak \textit{et al.} used the same real/imaginary character of the
orbital responses to remove the two-electron term from the MCSCF
AAT.\cite{Bak1993,Bak1994} If both perturbations have the same parity, such as the
DBOC for which both $x$ and $y$ correspond to nuclear coordinates, then the
two-electron contribution is non-vanishing, which is consistent with previous
formulations.\cite{Sellers:1984,Gauss2006} The elimination of the
two-electron-density contribution for mixed-parity perturbations represents a
substantial reduction in complexity for both the AAT and VG-APT relative to
previous work.  The present expression eliminates five terms for MP2 and 12
terms for CISD compared to our previous formulation.\cite{Shumberger2026jpca}
It should be noted, however, that the computational cost savings provided by
the density-based formalism do not imply that the VG-APT is less
computationally demanding than its LG-APT counterpart, as discussed in the
next section.

\section{Computational Details}
\label{sec:compdet}

The density-based CISD and MP2 AATs and VG-APTs (along with analytic Hessians
and LG-APTs) have been implemented in the PyCC package,\cite{pycc} which uses
\textsc{Psi4}\cite{Psi4} for its input Hamiltonian integrals, derivative
integrals, and SCF MO coefficients.  The implementation builds on an existing
HF-VCD layer that provides the CPHF solver and orbital response matrices for
nuclear and magnetic-field perturbations, while specialized MP2 and CI layers
compute the corresponding ground-state $c_I$ amplitudes, the derivative
$c_I^x$ amplitudes, and the necessary $D^{(0,0)}$, $D^{(x,0)}$, and
$D^{(0,y)}$ densities.  [Note that a fully normalized form of the first-order (MP1) wave
function is used for the MP2 implementation.] The computational infrastructure
for the Hessians and LG-APTs, on the other hand, follows well-established
energy-derivative formulations.\cite{Yamaguchi94, Handy85:MP2Hessians}

The computational costs of the AATs and VG-APTs are identical.  Both require
solution to the perturbed-amplitude equations with respect to $3N$ nuclear
coordinates as well as with respect to three magnetic-field (AAT) or
linear-momentum (VG-APT) components. However, given that these tensors are
normally computed as part of an overall VCD and/or IR spectrum simulation,
they are usually preceded by computation of the nuclear Hessian, and the
amplitude derivatives with respect to nuclear positions are normally
immediately available on disk.  Thus, the computation of the AAT and VG-APT
requires solutions of perturbed amplitudes with respect to
only three additional field/momentum perturbations, which is the same as required
for the LG-APT under the assumption that the latter is obtained based on
differentiation of the nuclear gradient with respect to the external electric
field.  In that case, the lack of two-electron-density-based terms in the
AAT/VG-APT results in a slightly less expensive final tensor assembly than in
the LG-APT case.  If, however, one chooses to compute the LG-APT as the
derivative of the relaxed electric dipole with respect to nuclear coordinates,
then no additional perturbed amplitudes are required, and the tensor assembly
is essentially free compared to other properties.  PyCC automatically stores
all derivative four-index tensor quantities on external disk once they are
computed in order to avoid re-solving any such expensive equations.

To verify the density-matrix formulation, we compared the AAT for \ce{H2O2}
for both MP2 and CISD using the STO-3G basis against the determinant-based
implementations of Refs.~\onlinecite{Shumberger2025jctc} and
\onlinecite{Shumberger2026jpca} and found agreement to within $10^{-10}$ a.u.\
across all elements. Thus, the density-matrix reformulation is algebraically
equivalent to the explicit determinant contraction while eliminating the
five (MP2) and 12 (CISD) two-electron density contributions in the
orbital-orbital block.

Basis set convergence was assessed across cc-pVDZ, cc-pVTZ, and
cc-pVQZ,\cite{Dunning89,Woon93} as well as aug-cc-pVDZ, aug-cc-pVTZ, and
aug-cc-pVQZ\cite{Kendall92,Woon93} for (\textit{P})-\ce{H2O2}, and the full
VCD pipeline was applied for both MP2 and CISD to \smo\ and \cby\ with the
cc-pVDZ, aug-cc-pVDZ, and cc-pVTZ basis sets.  Core orbitals ($1s$ for carbon
and oxygen, $1s2s2p$ for chlorine) were held frozen in all correlated wave
function calculations.  Optimized geometries for all molecules were obtained
using CFOUR\cite{Matthews2020:CFOUR} for MP2 and Gaussian 16\cite{Gaussian16} for CISD.

The present implementation does not include gauge-including atomic orbitals
(GIAOs) when computing the AATs.  However, origin independence of the VCD
rotatory strengths is obtained using either the VG-APT for the electric-dipole
transition moment or the recently reported length-gauge origin-independent
[LG(OI)] approach.\cite{Caricato2020, Caricato2021, Niemeyer2022,
Shumberger2026:LGOI} In Appendix \ref{app:lgoi}, we provide a simplified
reformulation of the LG(OI) for VCD rotatory strengths (that also applies to
most electronic CD rotatory strengths).  The molecular center of mass was used
for the coordinate origin for LG rotatory-strength calculations.

\section{Results}

\subsection{Basis Set Convergence}
\label{sec:basis}

\begin{figure*}[t]
\centering
\includegraphics[width=\textwidth]{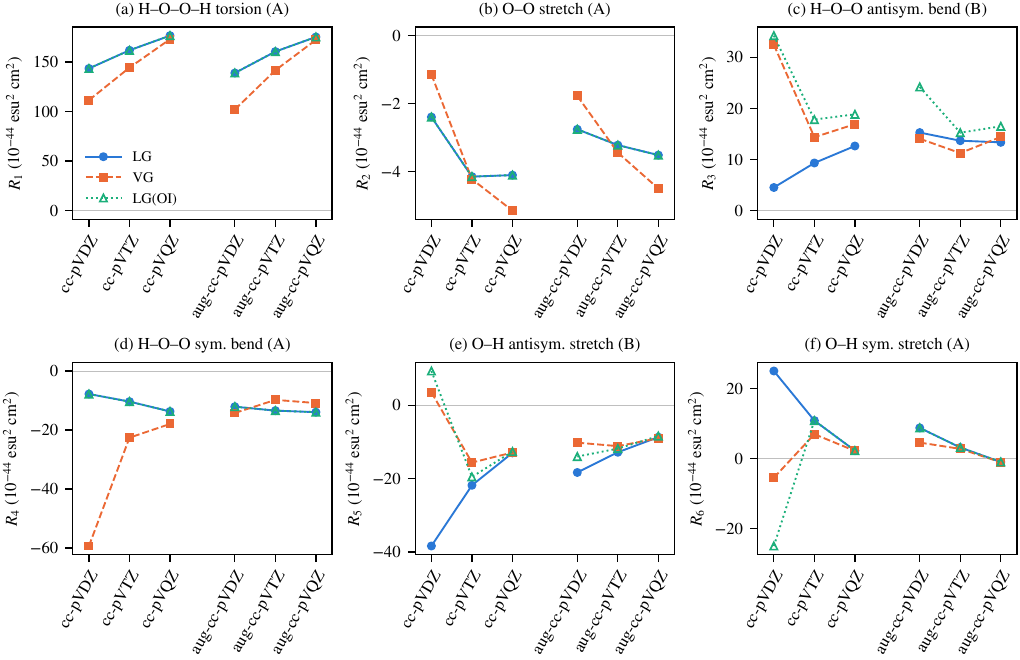}
\caption{Convergence of the MP2 VCD LG, VG, and LG(OI) rotatory strengths with
respect to basis set for each of the six vibrational modes of (\textit{P})-\ce{H2O2}.}
\label{fig:h2o2_mp2_convergence}
\end{figure*}

\begin{figure*}[t]
\centering
\includegraphics[width=\textwidth]{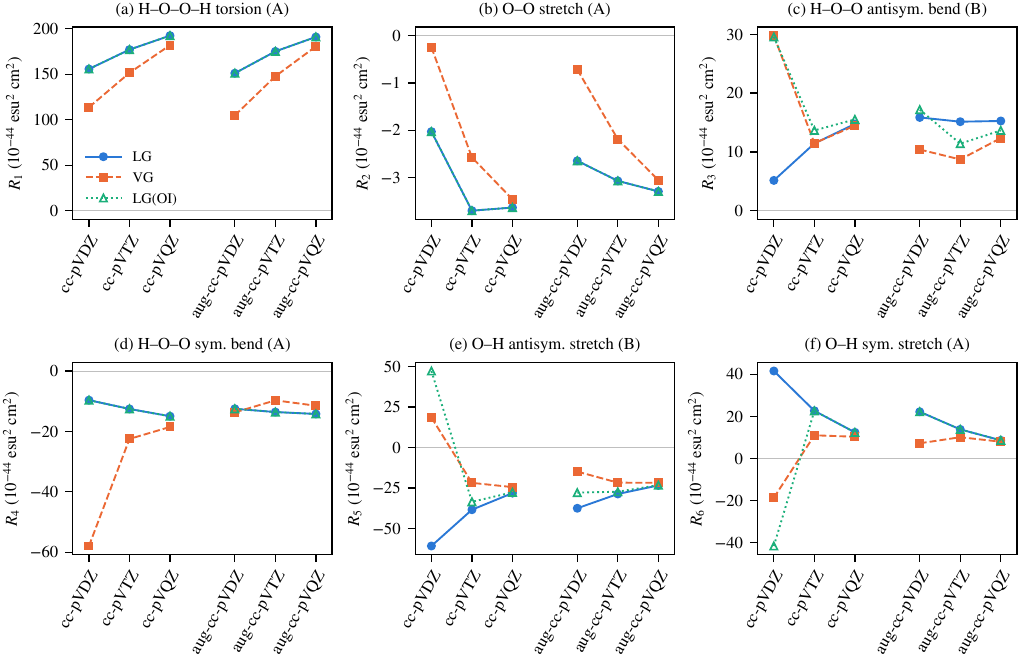}
\caption{Convergence of the CISD VCD LG, VG, and LG(OI) rotatory strengths with
respect to basis set for each of the six vibrational modes of (\textit{P})-\ce{H2O2}.}
\label{fig:h2o2_cisd_convergence}
\end{figure*}

Figs.~\ref{fig:h2o2_mp2_convergence} and \ref{fig:h2o2_cisd_convergence} plot
the convergence of the LG, VG, and LG(OI) VCD rotatory strengths with respect
to basis set for each of the six vibrational modes of (\textit{P})-\ce{H2O2} for MP2 and
CISD, respectively.  (Numerical values are provided in Tables
\ref{tab:si_h2o2_mp2} and \ref{tab:si_h2o2_cisd} of the SI.) The convergence
behavior of both methods is similar, with the DZ-type basis sets showing the
largest deviations, as expected.  With the cc-pVDZ basis, for example, the RMS
for LG and LG(OI) rotatory strengths relative to VG is 34.50 and 26.13 \runit,
respectively, for MP2 (49.37 and 30.15 for CISD), indicating that the basis
set is too far from complete to provide reliable results, even for a small
molecule such as \ce{H2O2}. For the aug-cc-pVQZ basis, however, the spread
between gauges is substantially reduced to 1.83 and 1.99 \runit\ for MP2 (4.60
and 4.47 for CISD), though the spread need not converge to zero for these
non-fully-variational methods, as noted in Appendix \ref{app:lgoi}.

It is noteworthy that, for all four $A$-irrep modes, the absolute values of
the LG and LG(OI) rotatory strengths are identical, regardless of method or
basis set. This is because the only non-zero components of the normal-mode LG-
and VG-APT for those modes are along the $z$-axis, which transforms as the $A$
irrep in the $C_2$ point group.  Thus, the LG- and VG-APTs are either parallel
or antiparallel to each other, and, as shown in Appendix \ref{app:lgoi}, the
LG(OI) rotatory strength equals the LG rotatory strength to within a sign.
Furthermore, in such cases the LG rotatory strength is naturally origin
invariant, as can be seen from Eq.~\eqref{eq:lg_origin_dependence}. For the
cc-pVDZ basis set for both MP2 and CISD, the VG-APT of mode 6 (the symmetric
O$-$H stretch) is antiparallel to the LG-APT, which explains the sign
discrepancy visible in the lower-right panel of both Figs.
\ref{fig:h2o2_mp2_convergence} and \ref{fig:h2o2_cisd_convergence}.  For the
other three $A$-irrep modes with that basis set (and all four modes for the
other basis sets), the LG and LG(OI) rotatory strengths overlap exactly in the
figures.

\subsection{Simulated VCD Spectra}
\label{sec:spectra}

\begin{figure*}[t]
\centering
\includegraphics[width=\textwidth]{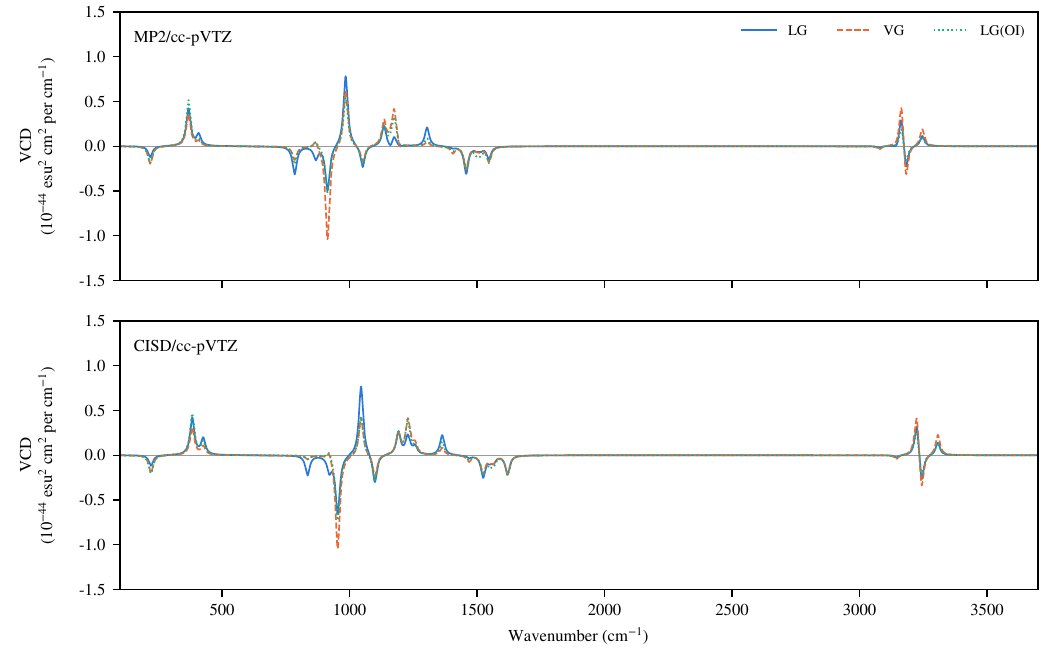}
\caption{MP2 (top) and CISD (bottom) VCD spectra of ($S$)-methyloxirane computed with the cc-pVTZ basis set,
showing the LG (blue, solid), VG (orange, dashed), and LG(OI) (green,
dotted) rotatory strengths. Each band is broadened with an area-normalized
Lorentzian function with a half-width at half-maximum of 10~cm$^{-1}$.}
\label{fig:smo_spectra}
\end{figure*}

\begin{figure*}[t]
\centering
\includegraphics[width=\textwidth]{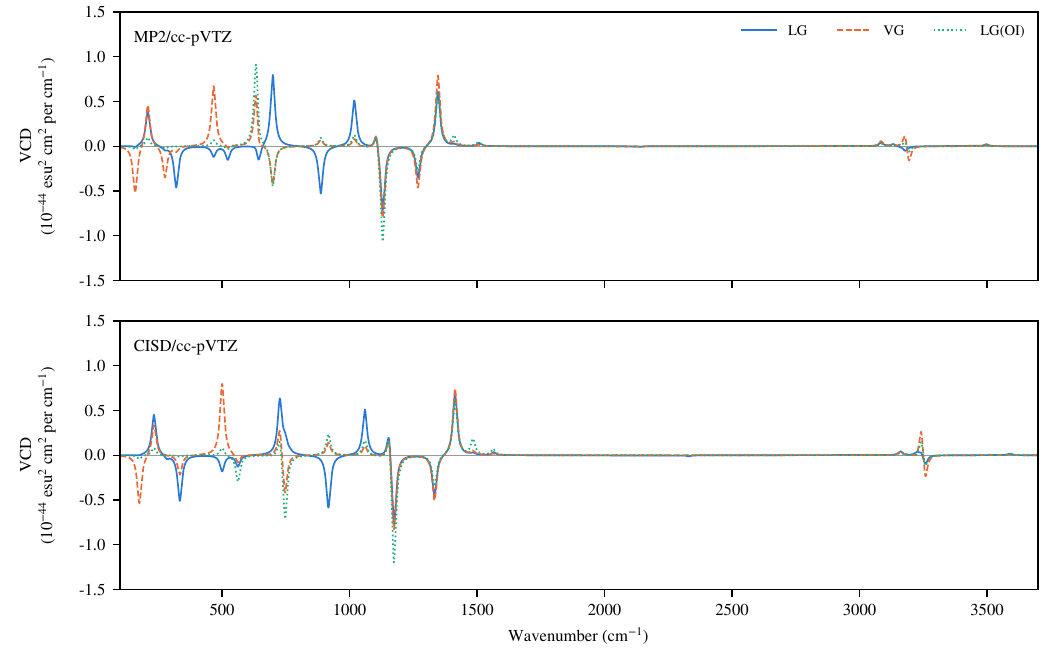}
\caption{MP2 (top) and CISD (bottom) VCD spectra of ($S$)-3-chloro-1-butyne computed with the cc-pVTZ basis set,
showing the LG (blue, solid), VG (orange, dashed), and LG(OI) (green,
dotted) rotatory strengths. Each band is broadened with an area-normalized
Lorentzian function with a half-width at half-maximum of 10~cm$^{-1}$.}
\label{fig:cby_spectra}
\end{figure*}

Having characterized the basis-set behavior for \ce{H2O2}, we now apply the
full VCD pipeline to two chiral molecules, \smo \ and \cby. Both are
single-conformer structures, and, while the former has served as a standard
test case for quantum chemical computations of a wide range of chiroptical
properties, the latter exhibits bands in the mid-IR that are well separated,
as noted by Polavarapu and co-workers.\cite{He2004} Both molecules were
treated at the MP2 and CISD levels with the cc-pVDZ, aug-cc-pVDZ, and cc-pVTZ
basis sets. The convergence behavior established in Sec.~\ref{sec:basis}
indicates that cc-pVTZ provides the most reliable results of the three (larger basis
sets are out of reach with our current implementation and computational
resources), so we present the cc-pVTZ results in the main text (Tables
\ref{tab:smo_vcd} and \ref{tab:cby_vcd} and Figs.~\ref{fig:smo_spectra} and
\ref{fig:cby_spectra}) and collect the cc-pVDZ and aug-cc-pVDZ results in the
SI. For each mode we report the LG, VG, and LG(OI) rotatory strengths together
with the cosine of the angle between the LG- and VG-APT vectors, $\cos\theta$,
a per-mode measure of LG/VG consistency.  

However, we note that even cc-pVTZ likely lacks the basis-set completeness
needed to represent the VG-APT accurately. Evidence for this is provided by
the fact that the APT summed over all nuclei must yield the net molecular
charge (the charge-sum rule\cite{Stephens1990}), which is zero for a neutral molecule.  The
LG-APTs satisfy this regardless of basis set, but the VG-APTs do not.  For
MP2, for example, the largest element of this sum for the \ce{H2O2} VG-APT
decreases consistently from 7.6 $e$ with the cc-pVDZ basis set down to 0.7 $e$
with the aug-cc-pVQZ basis set.  (A more complete set of values is given in
Table \ref{tab:si_sumrule} of the SI.) For \smo, the largest element of the
sum for the MP2 VG-APT decreases from 9.6 $e$ (cc-pVDZ) to 4.2 $e$ (cc-pVTZ),
and for \cby, the corresponding values are 21.8 $e$ and 13.3 $e$. CISD values
agree with MP2 to within 0.25 $e$.

For \smo, the three gauges are in close agreement at the cc-pVTZ level, where
the RMS differences of the LG and LG(OI) rotatory strengths relative to the VG
values are 4.51 and 2.98 \runit \ at the CISD level (5.14 and 4.39 for MP2).
At the CISD/cc-pVTZ level, the VCD spectrum is dominated by a strong negative
band at 953.6 cm$^{-1}$ ($R^{\rm VG} = -33.2$ \runit) with a weaker band of
opposite sign at 1045.2 cm$^{-1}$, together with a positive/negative couplet
in the C$-$H stretching region (3224.4 and 3243.5 cm$^{-1}$); the remaining
mid-infrared modes between 1200 and 1620 cm$^{-1}$ carry smaller rotatory
strengths. At the CISD level nearly all modes have $\cos\theta$ close to
unity and the three gauges agree quantitatively, with only three modes falling
below 0.8; the MP2 results are somewhat less consistent, with six such modes.
Only one CISD mode (two for MP2) shows a sign difference between the LG and VG
rotatory strengths, the modest band near 920 cm$^{-1}$ ($R^{\rm LG} = -5.2$,
$R^{\rm VG} = 2.9$ \runit); for this mode the LG(OI) value ($+2.6$ \runit)
follows the velocity-gauge sign, as the rank-one construction requires. The
other modes for which the gauges diverge, such as that at 1468.6 cm$^{-1}$
($\cos\theta = 0.56$), are of low intensity.

\begin{table*}[t]
\caption{MP2 and CISD VCD harmonic vibrational frequencies (cm$^{-1}$), LG, VG, and
LG(OI) rotatory strengths (10$^{-44}$ esu$^{2}$ cm$^{2}$), and the cosine of the angle
between the LG and VG APT vectors, $\cos\theta$, for each of the 24
vibrational modes of ($S$)-methyloxirane with the cc-pVTZ basis set. LG rotatory strengths were
computed with the molecular center of mass as the gauge origin. Modes are
numbered by ascending frequency at the CISD/cc-pVTZ level.}
\label{tab:smo_vcd}
\scriptsize
    \renewcommand{\arraystretch}{0.6}
    \setlength{\tabcolsep}{2pt}
    \begin{tabular*}{\textwidth}{@{\extracolsep{\fill}}lrrrrrcrrrrr}
    \toprule
     & \multicolumn{5}{c}{MP2/cc-pVTZ} & & \multicolumn{5}{c}{CISD/cc-pVTZ} \\
    \cmidrule{2-6} \cmidrule{8-12}
    Mode & \multicolumn{1}{c}{$\nu$} & \multicolumn{1}{c}{$R^{\rm LG}$} & \multicolumn{1}{c}{$R^{\rm VG}$} & \multicolumn{1}{c}{$R^{\rm LG(OI)}$} & \multicolumn{1}{c}{$\cos\theta$} & & \multicolumn{1}{c}{$\nu$} & \multicolumn{1}{c}{$R^{\rm LG}$} & \multicolumn{1}{c}{$R^{\rm VG}$} & \multicolumn{1}{c}{$R^{\rm LG(OI)}$} & \multicolumn{1}{c}{$\cos\theta$} \\ \midrule
    1 & 218.4 & $-$3.853 & $-$6.177 & $-$5.854 & 0.95 &  & 221.4 & $-$3.747 & $-$6.155 & $-$6.425 & 0.95 \\
    2 & 368.4 & 13.104 & 11.350 & 16.088 & 0.79 &  & 383.3 & 13.050 & 9.594 & 14.483 & 0.76 \\
    3 & 408.4 & 3.909 & 1.447 & 1.931 & 0.94 &  & 426.3 & 5.690 & 3.028 & 4.484 & 0.96 \\
    4 & 784.8 & $-$9.771 & $-$4.722 & $-$6.251 & 0.97 &  & 835.5 & $-$6.952 & $-$1.248 & $-$1.374 & 0.92 \\
    5 & 867.4 & $-$4.274 & 2.937 & 2.189 & 0.99 &  & 920.9 & $-$5.189 & 2.943 & 2.600 & 0.99 \\
    6 & 913.7 & $-$15.822 & $-$33.107 & $-$16.559 & 0.88 &  & 953.6 & $-$20.514 & $-$33.185 & $-$23.200 & 0.87 \\
    7 & 984.7 & 25.072 & 20.073 & 17.595 & 0.87 &  & 1045.2 & 24.683 & 13.705 & 14.141 & 0.94 \\
    8 & 1051.5 & $-$7.853 & $-$5.771 & $-$6.873 & 1.00 &  & 1099.3 & $-$10.242 & $-$7.802 & $-$9.756 & 1.00 \\
    9 & 1135.3 & 7.143 & 8.636 & 7.187 & 1.00 &  & 1191.9 & 7.944 & 7.458 & 7.664 & 1.00 \\
    10 & 1175.0\smash{$^{\rm a}$} & 3.456 & 12.464 & 9.833 & 0.91 &  & 1228.0 & 5.904 & 10.586 & 11.402 & 0.93 \\
    11 & 1161.0\smash{$^{\rm a}$} & $-$1.627 & 1.653 & 0.650 & $-$0.59 &  & 1236.6 & 0.968 & 2.730 & 1.012 & 0.91 \\
    12 & 1196.9 & $-$0.479 & $-$2.328 & $-$1.142 & 0.74 &  & 1257.2 & 2.817 & 3.397 & 2.414 & 0.99 \\
    13 & 1304.1 & 6.645 & 1.564 & 2.762 & 0.70 &  & 1363.4 & 7.085 & 2.650 & 4.662 & 0.86 \\
    14 & 1405.3 & $-$0.547 & $-$2.307 & $-$1.572 & 0.38 &  & 1468.6 & $-$0.836 & $-$2.147 & $-$1.490 & 0.56 \\
    15 & 1457.6 & $-$9.565 & $-$7.929 & $-$8.006 & 0.99 &  & 1524.2 & $-$7.578 & $-$6.321 & $-$6.357 & 0.99 \\
    16 & 1498.3 & $-$0.943 & $-$1.017 & $-$2.519 & 0.58 &  & 1550.4 & $-$1.566 & $-$1.500 & $-$2.873 & 0.72 \\
    17 & 1513.7 & $-$1.182 & $-$1.343 & $-$2.236 & 0.83 &  & 1565.1 & $-$2.105 & $-$2.033 & $-$2.759 & 0.85 \\
    18 & 1546.4 & $-$4.371 & $-$5.665 & $-$5.148 & 0.87 &  & 1620.0 & $-$6.633 & $-$6.716 & $-$6.841 & 0.95 \\
    19 & 3077.5 & $-$0.519 & $-$1.231 & $-$0.724 & 1.00 &  & 3144.4 & $-$0.870 & $-$1.458 & $-$1.007 & 1.00 \\
    20 & 3144.9 & $-$1.564 & $-$0.656 & $-$0.401 & 1.00 &  & 3215.8 & 1.730 & 2.780 & 1.957 & 1.00 \\
    21 & 3164.4\smash{$^{\rm a}$} & 9.151 & 16.475 & 7.404 & 1.00 &  & 3216.6 & $-$2.066 & $-$2.392 & $-$1.431 & 1.00 \\
    22 & 3162.1\smash{$^{\rm a}$} & 2.522 & 0.202 & 0.164 & 0.98 &  & 3224.4 & 12.557 & 15.670 & 9.652 & 0.97 \\
    23 & 3182.1 & $-$9.621 & $-$13.678 & $-$6.711 & 1.00 &  & 3243.5 & $-$11.014 & $-$13.994 & $-$8.955 & 1.00 \\
    24 & 3246.0 & 3.684 & 6.067 & 3.201 & 1.00 &  & 3307.1 & 4.441 & 7.258 & 4.487 & 1.00 \\
    \midrule
    \multicolumn{1}{l}{RMS LG$-$VG} & \multicolumn{5}{c}{5.14} &  & \multicolumn{5}{c}{4.51} \\
    \multicolumn{1}{l}{RMS VG$-$LG(OI)} & \multicolumn{5}{c}{4.39} &  & \multicolumn{5}{c}{2.98} \\
    \multicolumn{1}{l}{$N_{\rm sign}$ LG/VG} & \multicolumn{5}{c}{2} &  & \multicolumn{5}{c}{1} \\
    \bottomrule
    \end{tabular*}
    \par\vspace{2pt}\noindent\parbox{\textwidth}{\raggedright
    $^{\rm a}$Frequency order differs from the reference level (CISD/cc-pVTZ); the row follows the matched normal mode.
    $^{\rm b}$Normal-mode overlap with the reference mode $<0.80$.
    RMS: root-mean-square difference of the rotatory strengths over all modes. $N_{\rm sign}$: number of modes whose LG and VG rotatory strengths differ in sign, counted for modes with $\max(|R_1|,|R_2|)\geq1\times10^{-44}$ esu$^2$ cm$^2$.}

\end{table*}

For \cby, the spread between gauges is larger. At the CISD/cc-pVTZ level, the
RMS differences of the LG and LG(OI) rotatory strengths relative to VG are
10.48 and 7.28 \runit (12.55 and 6.95 for MP2), and the spread grows toward
smaller basis sets, reaching CISD RMS LG$-$VG values of 21.52 and 15.50 \runit
\ at cc-pVDZ and aug-cc-pVDZ, respectively (see SI). At the CISD/cc-pVTZ level
the VCD spectrum is dominated by the fingerprint region, with the strongest
band a negative feature at 1174.1 cm$^{-1}$ ($R^{\rm VG} = -28.2$ \runit) and
further strong bands at 1413.8 and 1333.3 cm$^{-1}$. The C$\equiv$C stretch at
2329.6 cm$^{-1}$ carries only a small rotatory strength, and the C$-$H
stretching region above 3100 cm$^{-1}$ includes the terminal alkyne C$-$H
stretch at 3588.7 cm$^{-1}$. The low-frequency skeletal modes below about 600
cm$^{-1}$ show the poorest gauge consistency: at the CISD/cc-pVTZ level four
modes have negative $\cos\theta$, ranging from $-0.27$ to $-0.93$, which we
again attribute primarily to incompleteness of the basis set. Only the mode
near 562 cm$^{-1}$ ($\cos\theta = -0.93$) approaches genuinely antiparallel
LG- and VG-APT vectors. The spectrally prominent band near 500 cm$^{-1}$ is in
fact the least antiparallel of the four ($\cos\theta = -0.27$); it is strong
in the velocity gauge ($R^{\rm VG} = 25.2$ \runit) but weak and of opposite
sign in the length gauge ($R^{\rm LG} = -5.6$ \runit), and because its
length-gauge dipole strength is much smaller than its velocity-gauge value,
its LG(OI) rotatory strength is also small ($R^{\rm LG(OI)} = 2.7$ \runit).
This is the converse of the LG(OI) magnification discussed below, where a
small velocity-gauge dipole strength instead inflates the LG(OI) value.

\begin{table*}[t]
\caption{MP2 and CISD VCD harmonic vibrational frequencies (cm$^{-1}$), LG, VG, and
LG(OI) rotatory strengths (10$^{-44}$ esu$^{2}$ cm$^{2}$), and the cosine of the angle
between the LG and VG APT vectors, $\cos\theta$, for each of the 24
vibrational modes of ($S$)-3-chloro-1-butyne with the cc-pVTZ basis set. LG rotatory strengths were
computed with the molecular center of mass as the gauge origin. Modes are
numbered by ascending frequency at the CISD/cc-pVTZ level.}
\label{tab:cby_vcd}
\scriptsize
    \renewcommand{\arraystretch}{0.6}
    \setlength{\tabcolsep}{2pt}
    \begin{tabular*}{\textwidth}{@{\extracolsep{\fill}}lrrrrrcrrrrr}
    \toprule
     & \multicolumn{5}{c}{MP2/cc-pVTZ} & & \multicolumn{5}{c}{CISD/cc-pVTZ} \\
    \cmidrule{2-6} \cmidrule{8-12}
    Mode & \multicolumn{1}{c}{$\nu$} & \multicolumn{1}{c}{$R^{\rm LG}$} & \multicolumn{1}{c}{$R^{\rm VG}$} & \multicolumn{1}{c}{$R^{\rm LG(OI)}$} & \multicolumn{1}{c}{$\cos\theta$} & & \multicolumn{1}{c}{$\nu$} & \multicolumn{1}{c}{$R^{\rm LG}$} & \multicolumn{1}{c}{$R^{\rm VG}$} & \multicolumn{1}{c}{$R^{\rm LG(OI)}$} & \multicolumn{1}{c}{$\cos\theta$} \\ \midrule
    1 & 159.2 & $-$0.668 & $-$16.875 & $-$1.285 & $-$0.73 &  & 175.9 & $-$0.332 & $-$17.149 & $-$1.385 & $-$0.74 \\
    2 & 208.9 & 12.402 & 15.258 & 3.093 & 0.11 &  & 232.8 & 14.394 & 11.052 & 2.645 & 0.10 \\
    3 & 276.4 & $-$1.033 & $-$11.102 & $-$0.905 & 0.12 &  & 282.8 & $-$1.344 & $-$0.167 & $-$0.017 & 0.23 \\
    4 & 320.4 & $-$14.535 & $-$1.694 & $-$0.139 & $-$0.81 &  & 334.6 & $-$16.101 & $-$6.947 & $-$0.602 & $-$0.76 \\
    5 & 467.1 & $-$3.580 & 21.124 & 2.100 & $-$0.43 &  & 500.3 & $-$5.621 & 25.174 & 2.741 & $-$0.27 \\
    6 & 522.6 & $-$4.682 & $-$1.229 & $-$1.270 & $-$0.91 &  & 562.3 & $-$3.925 & $-$3.125 & $-$9.300 & $-$0.93 \\
    7 & 699.0\smash{$^{\rm a}$} & 25.263 & $-$13.613 & $-$14.619 & $-$0.96 &  & 726.7 & 19.136 & 11.349 & 9.195 & $-$0.87 \\
    8 & 633.8\smash{$^{\rm a}$} & 1.140 & 21.447 & 32.064 & 0.97 &  & 747.0 & 4.368 & $-$15.347 & $-$24.169 & 0.97 \\
    9 & 643.6\smash{$^{\rm a}$} & $-$5.990 & $-$6.940 & $-$6.134 & 0.99 &  & 765.5 & 0.687 & 0.141 & 0.253 & 1.00 \\
    10 & 887.2 & $-$16.744 & 2.172 & 2.951 & 0.68 &  & 917.3 & $-$18.531 & 4.442 & 7.450 & 0.58 \\
    11 & 1018.7 & 16.278 & 2.671 & 4.130 & 0.43 &  & 1060.1 & 16.230 & 3.292 & 5.254 & 0.38 \\
    12 & 1105.4 & 5.690 & 6.947 & 8.061 & 0.95 &  & 1155.0 & 11.338 & 10.423 & 13.085 & 0.91 \\
    13 & 1130.0 & $-$23.132 & $-$26.012 & $-$34.605 & 0.86 &  & 1174.1 & $-$26.722 & $-$28.235 & $-$40.459 & 0.90 \\
    14 & 1268.1 & $-$11.306 & $-$14.812 & $-$8.818 & 0.98 &  & 1333.3 & $-$14.422 & $-$16.215 & $-$11.126 & 0.98 \\
    15 & 1346.6 & 19.315 & 25.092 & 19.082 & 0.99 &  & 1413.8 & 21.771 & 23.633 & 19.378 & 0.98 \\
    16 & 1409.5 & 0.546 & 1.188 & 3.515 & $-$0.49 &  & 1483.9 & 0.679 & 1.291 & 5.499 & 0.01 \\
    17 & 1500.6 & $-$0.502 & 0.235 & 0.457 & 0.97 &  & 1557.7 & $-$0.504 & $-$0.002 & $-$0.004 & 0.99 \\
    18 & 1506.7 & 1.494 & 0.289 & 0.884 & $-$0.22 &  & 1563.6 & 1.411 & 0.555 & 2.109 & 0.29 \\
    19 & 2138.1 & $-$0.247 & $-$0.110 & $-$0.270 & 1.00 &  & 2329.6 & $-$0.398 & $-$0.059 & $-$0.095 & 0.99 \\
    20 & 3084.4 & 1.354 & 1.791 & 0.959 & 1.00 &  & 3160.6 & 1.379 & 1.345 & 0.873 & 1.00 \\
    21 & 3130.8 & 0.832 & 0.374 & 0.147 & 0.95 &  & 3224.8 & 1.130 & $-$1.527 & $-$0.672 & 0.82 \\
    22 & 3179.2 & $-$1.588 & 5.516 & 2.198 & 1.00 &  & 3242.1 & 1.569 & 11.468 & 6.209 & 1.00 \\
    23 & 3193.4 & $-$0.195 & $-$6.733 & $-$2.548 & 1.00 &  & 3257.8 & $-$3.637 & $-$10.426 & $-$5.273 & 1.00 \\
    24 & 3497.2 & 0.739 & 0.431 & 0.530 & 1.00 &  & 3588.7 & 0.605 & 0.365 & 0.473 & 1.00 \\
    \midrule
    \multicolumn{1}{l}{RMS LG$-$VG} & \multicolumn{5}{c}{12.55} &  & \multicolumn{5}{c}{10.48} \\
    \multicolumn{1}{l}{RMS VG$-$LG(OI)} & \multicolumn{5}{c}{6.95} &  & \multicolumn{5}{c}{7.28} \\
    \multicolumn{1}{l}{$N_{\rm sign}$ LG/VG} & \multicolumn{5}{c}{4} &  & \multicolumn{5}{c}{4} \\
    \bottomrule
    \end{tabular*}
    \par\vspace{2pt}\noindent\parbox{\textwidth}{\raggedright
    $^{\rm a}$Frequency order differs from the reference level (CISD/cc-pVTZ); the row follows the matched normal mode.
    $^{\rm b}$Normal-mode overlap with the reference mode $<0.80$.
    RMS: root-mean-square difference of the rotatory strengths over all modes. $N_{\rm sign}$: number of modes whose LG and VG rotatory strengths differ in sign, counted for modes with $\max(|R_1|,|R_2|)\geq1\times10^{-44}$ esu$^2$ cm$^2$.}

\end{table*}

The MP2 and CISD spectra are qualitatively similar overall, but the degree of
agreement differs between the two molecules. Although the MP2 wave function
contains no singly excited determinants while CISD includes both singles and
doubles, the velocity-gauge rotatory strengths of \smo \ from the two methods
are well correlated (correlation coefficient 0.85, RMS difference 5.5 \runit)
and agree in sign for 14 of the 16 modes with $|R| \geq 3$ \runit. For \cby \
the agreement is weaker (correlation 0.67, RMS 9.8 \runit, sign agreement 14
of 17), with the largest method differences in the congested skeletal region
between 630 and 765 cm$^{-1}$, where the normal modes reorder between
the two methods. The CISD harmonic frequencies are uniformly higher than the
MP2 values, by 6.0\% on average at cc-pVTZ and by as much as 19\% for the
reordered modes near 765 cm$^{-1}$.

The LG(OI) rotatory strengths require closer attention, since the construction
is sensitive to weak modes. From the rank-one relation derived in Appendix
\ref{app:lgoi}, $R^{\rm LG(OI)} = R^{\rm VG}\sqrt{D^{\rm LG}/D^{\rm VG}}$, the
LG(OI) value is the VG value scaled by $\sqrt{D^{\rm LG}/D^{\rm VG}} = |P^{\rm
LG}|/|P^{\rm VG}|$, the ratio of the LG- and VG-APT magnitudes for that mode.
This ratio reflects how well the hypervirial relation holds: for modes whose
two gauges agree, it is close to unity and the three rotatory strengths track
one another, but for a mode with a small VG dipole strength, the ratio can
grow large and the LG(OI) value is magnified. Such modes recur across the
spectra, typically one or two per basis set, and are most pronounced for \cby
\ at the smaller basis sets (they are largely absent from the better-behaved
\smo \ spectra). The clearest case is the CISD/aug-cc-pVDZ mode at 721.7
cm$^{-1}$ (Table \ref{tab:si_cby_cisd} and Fig. \ref{fig:si_cby_apvdz} of the
SI), whose LG(OI) rotatory strength of 85.5 \runit \ far exceeds both its LG
(12.7) and VG (4.8) values. This mode has the second-smallest VG dipole
strength of the spectrum ($D^{\rm VG} = 1.2$, in units of $10^{-40}$ esu$^2$
cm$^2$), while its much larger LG dipole strength ($D^{\rm LG} = 389$) gives a
magnification factor of 17.8 and LG- and VG-APT vectors that are nearly
orthogonal ($\cos\theta = 0.17$). Milder magnification of the same origin
occurs at the CISD level for the mode at 1529 cm$^{-1}$ (aug-cc-pVDZ) and at
562 and 1484 cm$^{-1}$ (cc-pVTZ), among others. In every case the inflated
LG(OI) value reflects a small VG dipole strength rather than the AAT itself,
and the velocity-gauge rotatory strength, which is origin independent, is
unaffected. The magnification depends on the basis set but not monotonically:
for \cby \ it is largest at aug-cc-pVDZ rather than at the smaller cc-pVDZ.
Individual modes can nonetheless improve with basis, the 721.7 cm$^{-1}$ mode
being well-behaved at cc-pVTZ, where it appears at 765.5 cm$^{-1}$ with
$\cos\theta = 1.00$ and an LG(OI) rotatory strength of only 0.25 \runit.
Because CISD optimizes only the configuration coefficients and not the
orbitals, and MP2 is not variational at all, neither satisfies the hypervirial
relation, and the LG and VG values need not coincide even in the
complete-basis limit (Appendix \ref{app:lgoi}).

However, while the choice of method contributes to the differences, the main
source of discrepancies between LG- and VG-APTs (and thus rotatory strengths) is
incompleteness of the basis set.  Our preliminary studies indicate that this
is largely due to the contribution from the Hartree-Fock reference (as suggested
by the data in Table \ref{tab:si_sumrule}), but a detailed study of the basis
set convergence of the MP2 and CISD VCD rotatory strengths is outside the
scope of this work and will be treated in a subsequent paper.

\section{Conclusions}
\label{sec:conclusions}

We have reformulated the AAT and the VG-APT as contractions of perturbed and
unperturbed unrelaxed density matrices, valid for any real, normalized wave
function and applied here to MP2 and CISD. Following the elegant formulation
of Gauss and co-workers of the DBOC,\cite{Gauss2006} we determined that the
two-electron contribution to the orbital-orbital block vanishes by symmetry
because the real (nuclear) and imaginary (magnetic field or momentum)
perturbations have opposite parity, as they do for the AAT and the VG-APT, a
finding also noted by Bak \textit{et al.} for their MCSCF formulation of the
AAT.\cite{Bak1993,Bak1994} The two properties then reduce to only one-electron density
expressions, unlike the DBOC, which requires the unperturbed two-electron
density because both perturbations are real.  This formulation recovers the
known SCF-level expressions, but eliminates five terms for MP2 and 12 terms
for CISD from the explicit determinant contractions of our earlier
approach,\cite{Shumberger2026jpca} against which we have validated the new MP2
and CISD AATs.  In addition, the computational cost of the AAT and VG-APT is
the same, and, once the nuclear Hessian has been computed, each requires only
three additional perturbed-amplitude solutions.  The VG-APT yields natural
origin invariance for the rotatory strength, and we have also simplified the
LG(OI) approach for origin-invariant rotatory strengths that we reported in a
previous paper\cite{Shumberger2026:LGOI} to avoid the use of a singular-value
decomposition (SVD) of the outer product of the LG- and VG-APT vectors.

We have applied this new formulation to three chiral test molecules,
(\textit{P})-\ce{H2O2}, \smo, and \cby.  For \ce{H2O2}, we examined the
behavior of the LG, VG, and LG(OI) rotatory strengths across a series of basis
sets up to aug-cc-pVQZ, and determined that the VG rotatory strength converges
more slowly than its LG counterpart, as expected.  The RMS difference in
rotatory strengths across LG, VG, and LG(OI) decreases with increasing
basis-set size, but will not necessarily fall to zero in the
complete-basis-set limit because neither MP2 nor CISD satisfies the
hypervirial relation.  We have found that the charge-sum rule provides a
useful diagnostic for basis-set completeness that is independent of the
correlation model.  While the LG-APT obeys it exactly, the VG-APT does not,
but the violation decreases steadily with the size of the basis set.
Furthermore, preliminary data (see the SI) indicate that the bulk of the
charge-sum-rule discrepancy comes from the Hartree-Fock reference rather than
the correlated wave function.  For \smo\ and \cby, we find that the MP2 and
CISD VCD spectra are qualitatively similar, but the two exhibit much greater
agreement for the former than the latter.  For \cby, the largest differences
appear in the $630-765$ cm$^{-1}$ region, which includes the C$-$Cl stretch
and the two components of the acetylenic C$-$H bend.  These results motivate
the extension of this work to more comprehensive basis-set effects as well as
the use of gauge-including atomic orbitals.

\section{Supporting Information} \label{si}

Optimized geometries of all test molecules; harmonic vibrational frequencies, LG, VG, and mixed
dipole strengths, and LG, VG, and LG(OI) rotatory strengths of all test molecules with all basis
sets, together with $\cos\theta$ for ($S$)-methyloxirane and ($S$)-3-chloro-1-butyne; VCD
spectra of ($S$)-methyloxirane and ($S$)-3-chloro-1-butyne with the cc-pVDZ and aug-cc-pVDZ
basis sets; and the charge-sum rule for the VG-APT of all test molecules with all basis sets.

\section{Acknowledgements} \label{ack}

This work was supported by the U.S.\ National Science Foundation via grants
CHE-2154753 and OAC-2410880.  The authors are grateful to Advanced Research
Computing at Virginia Tech for providing computational resources that have
contributed to the results reported within the paper.

\appendix
\section{Reformulation of the LG(OI) Rotatory Strength}
\label{app:lgoi}
The VCD rotatory strength $R_i$ of a given vibrational mode, $i$, is the dot
product of the normal-mode representations of the APT,
$\vec{P}_i$, and the AAT, $\M$.  The APT may be computed
using the length gauge (LG) or velocity gauge (VG), and thus the rotatory
strength is dependent on the choice of gauge (omitting constant factors from this point forward),
\begin{equation}
\Rlg = \Plg \cdot \M \ \ \ \text{or} \ \ \ \Rvg = \Pvg \cdot \M.
\end{equation}
Similarly, the dipole strength ($D_i$) is the dot product of the normal-mode APTs, such that
\begin{equation}
\begin{gathered}
\Dlg = \Plg \cdot \Plg \ \ \ \text{or} \ \ \ \Dvg = \Pvg \cdot \Pvg \\
\text{or} \ \ \ \Dmix = \Plg \cdot \Pvg.
\end{gathered}
\end{equation}
While the APTs are origin invariant, the LG
rotatory strength, $\Rlg$, is not; translation of the coordinate origin along a
vector $\vec{B}$ shifts the rotatory strength by an amount that depends on the
cross product of the LG- and VG-APTs, \textit{viz.},
\begin{equation}
\Rlg \leftarrow \Rlg + \vec{B} \cdot (\Plg \times \Pvg).
\label{eq:lg_origin_dependence}
\end{equation}
The VG rotatory strength is origin invariant, however, 
\begin{equation}
\Rvg \leftarrow \Rvg + \vec{B} \cdot (\Pvg \times \Pvg) = \Rvg
\end{equation}
because the cross product of $\Pvg$ with itself is zero. In the limit of a complete basis set, $\Plg$
and $\Pvg$ are the same for exact wave functions and for fully variational methods (\textit{e.g.}, HF, 
MCSCF, or Kohn-Sham DFT), in which case $\Rlg$ and $\Rvg$ become identical.

In the LG(OI) approach,\cite{Shumberger2026:LGOI} the origin dependence of $\Rlg$ is removed by
computing the SVD of the outer product of the LG- and VG-APTs,
\begin{equation}
\Di = \Plg (\Pvg)^\dagger = \bm{U}_i \bm{\Sigma}_i \bm{V}_i^\dagger,
\end{equation}
so that the cross product of the transformed $\bm{U}_i^\dagger \Plg$ and $\bm{V}_i^\dagger \Pvg$ vectors
is zero.  As a result, the transformation of the LG rotatory strength,
\begin{equation}
\Rlgoi = \text{tr}(\bm{U}_i^\dagger \Ri \bm{V}_i),
\label{eq:lgoi}
\end{equation}
where $\Ri = \Plg (\M)^\dagger$, is origin invariant.

However, a simpler approach to computing the LG(OI) rotatory strengths may be obtained by recognizing
that the matrix $\Di$ is rank one because its columns may be written as the vector $\Plg$ scaled
by the components of $\Pvg$ and its rows are the vector $\Pvg$ scaled by the components of $\Plg$,
that is,
\begin{align}
\Di &= \Plg (\Pvg)^\dagger \nonumber \\
&= \begin{pmatrix}
(\Pvg)_1 \Plg & (\Pvg)_2 \Plg & (\Pvg)_3 \Plg \end{pmatrix} \nonumber \\
        &= \begin{pmatrix} (\Plg)_1 (\Pvg)^\dagger \\ (\Plg)_2 (\Pvg)^\dagger \\ (\Plg)_3 (\Pvg)^\dagger \end{pmatrix}.
\end{align}
Thus, the column space of $\Di$ is spanned by $\Plg$ alone and its row space is spanned by $\Pvg$ alone,
yielding only one non-zero singular value corresponding to the first column of the left and right
transformation matrices, respectively,
\begin{equation}
(\vec{u}_i)_1 = s\frac{\Plg}{|\Plg|} \ \ \ \text{and} \ \ \ (\vec{v}_i)_1 = s\frac{\Pvg}{|\Pvg|},
\end{equation}
where $s = \pm 1$.
Rewriting the trace in Eq.~\eqref{eq:lgoi} explicitly and recognizing that the columns of $\bm{U}_i$ are
orthonormal, we find that
\begin{align}
\Rlgoi &= \sum_{k=1}^3 (\vec{u}_i)_k^\dagger \Plg (\M)^\dagger (\vec{v}_i)_k \nonumber \\
&= \sum_{k=1}^3 \left[(\vec{u}_i)_k \cdot \Plg \right] \left[\M \cdot (\vec{v}_i)_k\right] \nonumber \\
&= \left[(\vec{u}_i)_1 \cdot \Plg \right] \left[\M \cdot (\vec{v}_i)_1\right] \nonumber \\
&= \left( \frac{\Plg \cdot \Plg}{|\Plg|} \right) \left( \frac{\M \cdot \Pvg}{|\Pvg|} \right) \nonumber \\
&= \Rvg \sqrt{ \frac{\Dlg}{\Dvg} }.
\label{eq:new_lgoi}
\end{align}
Thus, the SVD of $\Di$ is not necessary because $\Rlgoi$ derives its origin invariance from that of
$\Rvg$, $\Dvg$, and $\Dlg$.  In addition, the sign of $\Rlgoi$ always matches that of $\Rvg$, and if
the two APT vectors are parallel or antiparallel, $\Rlgoi = \text{sgn}(\Dmix) \Rlg$.

While Eq.~\eqref{eq:new_lgoi} is specific to VCD rotatory strengths, the same
concept applies to the LG(OI) formulation of electronic CD rotatory strengths developed by
Niemeyer, Caricato, and Neugebauer,\cite{Niemeyer2022} where the mixed-gauge
electric-dipole transition-strength tensor for a given electronic transition
(the analogue of $\Di$) is also rank one (except for methods with distinct left
and right transition moments, such as linear-response coupled cluster, where it may be rank two).  The
same is not true, however, of the LG(OI) formulation of optical rotation by
Caricato\cite{Caricato2020} and by Caricato and Balduf\cite{Caricato2021}
because the mixed-gauge electric-dipole polarizability $\alpha^{(R,P)}(\omega)$
is generally full rank.

\bibliography{references}
\end{document}


\maketitle
\tableofcontents
\clearpage


\section{Optimized geometries}
\label{si:geometries}

Cartesian coordinates (bohr) in the center-of-mass frame.

\small
\subsection{($P$)-H$_2$O$_2$}
\noindent
\begin{minipage}[t]{0.49\textwidth}\centering
\setlength{\tabcolsep}{4pt}
\begin{tabular}{lrrr}
\toprule
\multicolumn{4}{c}{MP2/cc-pVDZ} \\
\midrule
Atom & \multicolumn{1}{c}{$x$} & \multicolumn{1}{c}{$y$} & \multicolumn{1}{c}{$z$} \\
\midrule
H & $-$1.7813513242 & 1.4124614608 & 0.8724949210 \\
O & $-$1.3715896179 & $-$0.1155859947 & $-$0.0549751120 \\
O & 1.3715896179 & 0.1155859947 & $-$0.0549751120 \\
H & 1.7813513242 & $-$1.4124614608 & 0.8724949210 \\
\bottomrule
\end{tabular}
\end{minipage}%
\hfill
\begin{minipage}[t]{0.49\textwidth}\centering
\setlength{\tabcolsep}{4pt}
\begin{tabular}{lrrr}
\toprule
\multicolumn{4}{c}{MP2/cc-pVTZ} \\
\midrule
Atom & \multicolumn{1}{c}{$x$} & \multicolumn{1}{c}{$y$} & \multicolumn{1}{c}{$z$} \\
\midrule
H & $-$1.7841162937 & 1.3684187797 & 0.9182133677 \\
O & $-$1.3658645952 & $-$0.1126257648 & $-$0.0578557897 \\
O & 1.3658645952 & 0.1126257648 & $-$0.0578557897 \\
H & 1.7841162937 & $-$1.3684187797 & 0.9182133677 \\
\bottomrule
\end{tabular}
\end{minipage}%
\par\bigskip
\noindent
\begin{minipage}[t]{0.49\textwidth}\centering
\setlength{\tabcolsep}{4pt}
\begin{tabular}{lrrr}
\toprule
\multicolumn{4}{c}{MP2/cc-pVQZ} \\
\midrule
Atom & \multicolumn{1}{c}{$x$} & \multicolumn{1}{c}{$y$} & \multicolumn{1}{c}{$z$} \\
\midrule
H & $-$1.7886492601 & 1.3518922242 & 0.9349748778 \\
O & $-$1.3620727173 & $-$0.1118588060 & $-$0.0589119172 \\
O & 1.3620727173 & 0.1118588060 & $-$0.0589119172 \\
H & 1.7886492601 & $-$1.3518922242 & 0.9349748778 \\
\bottomrule
\end{tabular}
\end{minipage}%
\hfill
\begin{minipage}[t]{0.49\textwidth}\centering
\setlength{\tabcolsep}{4pt}
\begin{tabular}{lrrr}
\toprule
\multicolumn{4}{c}{MP2/aug-cc-pVDZ} \\
\midrule
Atom & \multicolumn{1}{c}{$x$} & \multicolumn{1}{c}{$y$} & \multicolumn{1}{c}{$z$} \\
\midrule
H & $-$1.7974012806 & 1.3704014439 & 0.9422257298 \\
O & $-$1.3846433914 & $-$0.1120877452 & $-$0.0593687868 \\
O & 1.3846433914 & 0.1120877452 & $-$0.0593687868 \\
H & 1.7974012806 & $-$1.3704014439 & 0.9422257298 \\
\bottomrule
\end{tabular}
\end{minipage}%
\par\bigskip
\noindent
\begin{minipage}[t]{0.49\textwidth}\centering
\setlength{\tabcolsep}{4pt}
\begin{tabular}{lrrr}
\toprule
\multicolumn{4}{c}{MP2/aug-cc-pVTZ} \\
\midrule
Atom & \multicolumn{1}{c}{$x$} & \multicolumn{1}{c}{$y$} & \multicolumn{1}{c}{$z$} \\
\midrule
H & $-$1.7952794871 & 1.3563765901 & 0.9405975201 \\
O & $-$1.3689679195 & $-$0.1120784972 & $-$0.0592661948 \\
O & 1.3689679195 & 0.1120784972 & $-$0.0592661948 \\
H & 1.7952794871 & $-$1.3563765901 & 0.9405975201 \\
\bottomrule
\end{tabular}
\end{minipage}%
\hfill
\begin{minipage}[t]{0.49\textwidth}\centering
\setlength{\tabcolsep}{4pt}
\begin{tabular}{lrrr}
\toprule
\multicolumn{4}{c}{MP2/aug-cc-pVQZ} \\
\midrule
Atom & \multicolumn{1}{c}{$x$} & \multicolumn{1}{c}{$y$} & \multicolumn{1}{c}{$z$} \\
\midrule
H & $-$1.7943270834 & 1.3517365633 & 0.9362359083 \\
O & $-$1.3626606404 & $-$0.1121525565 & $-$0.0589913736 \\
O & 1.3626606404 & 0.1121525565 & $-$0.0589913736 \\
H & 1.7943270834 & $-$1.3517365633 & 0.9362359083 \\
\bottomrule
\end{tabular}
\end{minipage}%
\par\bigskip
\noindent
\begin{minipage}[t]{0.49\textwidth}\centering
\setlength{\tabcolsep}{4pt}
\begin{tabular}{lrrr}
\toprule
\multicolumn{4}{c}{CISD/cc-pVDZ} \\
\midrule
Atom & \multicolumn{1}{c}{$x$} & \multicolumn{1}{c}{$y$} & \multicolumn{1}{c}{$z$} \\
\midrule
H & $-$1.7806223500 & 1.3966216600 & 0.8947991536 \\
O & $-$1.3507700600 & $-$0.0938845296 & $-$0.0563804814 \\
O & 1.3507700600 & 0.0938845296 & $-$0.0563804814 \\
H & 1.7806223500 & $-$1.3966216600 & 0.8947991536 \\
\bottomrule
\end{tabular}
\end{minipage}%
\hfill
\begin{minipage}[t]{0.49\textwidth}\centering
\setlength{\tabcolsep}{4pt}
\begin{tabular}{lrrr}
\toprule
\multicolumn{4}{c}{CISD/cc-pVTZ} \\
\midrule
Atom & \multicolumn{1}{c}{$x$} & \multicolumn{1}{c}{$y$} & \multicolumn{1}{c}{$z$} \\
\midrule
H & $-$1.7833397300 & 1.3530028900 & 0.9250773344 \\
O & $-$1.3380575300 & $-$0.0908112149 & $-$0.0582882820 \\
O & 1.3380575300 & 0.0908112149 & $-$0.0582882820 \\
H & 1.7833397300 & $-$1.3530028900 & 0.9250773344 \\
\bottomrule
\end{tabular}
\end{minipage}%
\par\bigskip
\noindent
\begin{minipage}[t]{0.49\textwidth}\centering
\setlength{\tabcolsep}{4pt}
\begin{tabular}{lrrr}
\toprule
\multicolumn{4}{c}{CISD/cc-pVQZ} \\
\midrule
Atom & \multicolumn{1}{c}{$x$} & \multicolumn{1}{c}{$y$} & \multicolumn{1}{c}{$z$} \\
\midrule
H & $-$1.7875507100 & 1.3375247300 & 0.9343419017 \\
O & $-$1.3316122900 & $-$0.0913923235 & $-$0.0588720340 \\
O & 1.3316122900 & 0.0913923235 & $-$0.0588720340 \\
H & 1.7875507100 & $-$1.3375247300 & 0.9343419017 \\
\bottomrule
\end{tabular}
\end{minipage}%
\hfill
\begin{minipage}[t]{0.49\textwidth}\centering
\setlength{\tabcolsep}{4pt}
\begin{tabular}{lrrr}
\toprule
\multicolumn{4}{c}{CISD/aug-cc-pVDZ} \\
\midrule
Atom & \multicolumn{1}{c}{$x$} & \multicolumn{1}{c}{$y$} & \multicolumn{1}{c}{$z$} \\
\midrule
H & 1.3792236700 & 1.7776321700 & 0.9437908136 \\
O & $-$0.0749102181 & 1.3549118200 & $-$0.0594674014 \\
O & 0.0749102182 & $-$1.3549118200 & $-$0.0594674012 \\
H & $-$1.3792236700 & $-$1.7776321700 & 0.9437908136 \\
\bottomrule
\end{tabular}
\end{minipage}%
\par\bigskip
\noindent
\begin{minipage}[t]{0.49\textwidth}\centering
\setlength{\tabcolsep}{4pt}
\begin{tabular}{lrrr}
\toprule
\multicolumn{4}{c}{CISD/aug-cc-pVTZ} \\
\midrule
Atom & \multicolumn{1}{c}{$x$} & \multicolumn{1}{c}{$y$} & \multicolumn{1}{c}{$z$} \\
\midrule
H & $-$1.7862325700 & 1.3509160800 & 0.9387041546 \\
O & $-$1.3384936700 & $-$0.0852398139 & $-$0.0591468956 \\
O & 1.3384936700 & 0.0852398139 & $-$0.0591468956 \\
H & 1.7862325700 & $-$1.3509160800 & 0.9387041546 \\
\bottomrule
\end{tabular}
\end{minipage}%
\hfill
\begin{minipage}[t]{0.49\textwidth}\centering
\setlength{\tabcolsep}{4pt}
\begin{tabular}{lrrr}
\toprule
\multicolumn{4}{c}{CISD/aug-cc-pVQZ} \\
\midrule
Atom & \multicolumn{1}{c}{$x$} & \multicolumn{1}{c}{$y$} & \multicolumn{1}{c}{$z$} \\
\midrule
H & $-$1.7903119300 & 1.3389985900 & 0.9341585783 \\
O & $-$1.3313971300 & $-$0.0905629930 & $-$0.0588604830 \\
O & 1.3313971300 & 0.0905629930 & $-$0.0588604830 \\
H & 1.7903119300 & $-$1.3389985900 & 0.9341585783 \\
\bottomrule
\end{tabular}
\end{minipage}%
\par\bigskip
\subsection{($S$)-Methyloxirane}
\noindent
\begin{minipage}[t]{0.49\textwidth}\centering
\setlength{\tabcolsep}{4pt}
\begin{tabular}{lrrr}
\toprule
\multicolumn{4}{c}{MP2/cc-pVDZ} \\
\midrule
Atom & \multicolumn{1}{c}{$x$} & \multicolumn{1}{c}{$y$} & \multicolumn{1}{c}{$z$} \\
\midrule
C & $-$0.4004485637 & $-$0.0835383304 & 0.9321659815 \\
C & 1.8207946270 & $-$1.3081730619 & $-$0.2202542557 \\
O & 1.5369563023 & 1.3959348459 & $-$0.2695974464 \\
C & $-$2.9445432019 & $-$0.0951542212 & $-$0.3454637527 \\
H & $-$0.4174884526 & 0.0957657469 & 3.0022564638 \\
H & 1.6058429558 & $-$2.2528381250 & $-$2.0533697448 \\
H & 3.3497227304 & $-$2.0172704708 & 0.9857805933 \\
H & $-$2.7077576385 & $-$0.2711367254 & $-$2.4058428544 \\
H & $-$3.9670467445 & 1.6777997115 & 0.0450612576 \\
H & $-$4.1075304997 & $-$1.6829650098 & 0.3415881652 \\
\bottomrule
\end{tabular}
\end{minipage}%
\hfill
\begin{minipage}[t]{0.49\textwidth}\centering
\setlength{\tabcolsep}{4pt}
\begin{tabular}{lrrr}
\toprule
\multicolumn{4}{c}{MP2/aug-cc-pVDZ} \\
\midrule
Atom & \multicolumn{1}{c}{$x$} & \multicolumn{1}{c}{$y$} & \multicolumn{1}{c}{$z$} \\
\midrule
C & $-$0.4071811013 & $-$0.1040629777 & 0.9333032070 \\
C & 1.8118388759 & $-$1.3290821287 & $-$0.2274844055 \\
O & 1.5530746076 & 1.4139944860 & $-$0.2597623345 \\
C & $-$2.9497489061 & $-$0.0816953912 & $-$0.3488012902 \\
H & $-$0.4134509272 & 0.0753095061 & 2.9984885583 \\
H & 1.6025005182 & $-$2.2269944334 & $-$2.0794637899 \\
H & 3.3475450239 & $-$2.0297095818 & 0.9656863929 \\
H & $-$2.7171678827 & $-$0.2762105121 & $-$2.4055031419 \\
H & $-$3.9442756440 & 1.7043891479 & 0.0397327084 \\
H & $-$4.1264367444 & $-$1.6509564413 & 0.3527296813 \\
\bottomrule
\end{tabular}
\end{minipage}%
\par\bigskip
\noindent
\begin{minipage}[t]{0.49\textwidth}\centering
\setlength{\tabcolsep}{4pt}
\begin{tabular}{lrrr}
\toprule
\multicolumn{4}{c}{MP2/cc-pVTZ} \\
\midrule
Atom & \multicolumn{1}{c}{$x$} & \multicolumn{1}{c}{$y$} & \multicolumn{1}{c}{$z$} \\
\midrule
C & 0.3799360184 & 0.0056032720 & 0.9405113160 \\
C & $-$1.8521439315 & 1.2463755145 & $-$0.1156571571 \\
O & $-$1.4665303016 & $-$1.4206455932 & $-$0.4420176428 \\
C & 2.9222702161 & 0.2509006060 & $-$0.2760047276 \\
H & 0.3667723266 & $-$0.3907220683 & 2.9515765859 \\
H & $-$1.6421178124 & 2.3600251722 & $-$1.8193353422 \\
H & $-$3.4135730689 & 1.7499981040 & 1.1046744252 \\
H & 2.7063001868 & 0.6363068332 & $-$2.2838085955 \\
H & 4.0001702887 & $-$1.4880715212 & $-$0.0591693560 \\
H & 3.9917047767 & 1.7846125222 & 0.5861468859 \\
\bottomrule
\end{tabular}
\end{minipage}%
\hfill
\begin{minipage}[t]{0.49\textwidth}\centering
\setlength{\tabcolsep}{4pt}
\begin{tabular}{lrrr}
\toprule
\multicolumn{4}{c}{CISD/cc-pVDZ} \\
\midrule
Atom & \multicolumn{1}{c}{$x$} & \multicolumn{1}{c}{$y$} & \multicolumn{1}{c}{$z$} \\
\midrule
C & 0.6069950483 & $-$0.4535168762 & 0.6596716772 \\
C & 0.6130242723 & $-$0.4544668362 & $-$2.1109848138 \\
O & 0.5984129053 & 1.8414956588 & $-$0.7247196348 \\
C & $-$1.7347805267 & $-$1.0330602812 & 2.1647482762 \\
H & 2.3998012033 & $-$0.8413475182 & 1.6103819562 \\
H & $-$1.1309531067 & $-$0.8865055912 & $-$3.1235260038 \\
H & 2.3547559533 & $-$0.8578811492 & $-$3.1373346438 \\
H & $-$3.4292832067 & $-$0.5775873152 & 1.0636956262 \\
H & $-$1.7717781567 & 0.0895441175 & 3.9080944662 \\
H & $-$1.7906162567 & $-$3.0404166912 & 2.6857651362 \\
\bottomrule
\end{tabular}
\end{minipage}%
\par\bigskip
\noindent
\begin{minipage}[t]{0.49\textwidth}\centering
\setlength{\tabcolsep}{4pt}
\begin{tabular}{lrrr}
\toprule
\multicolumn{4}{c}{CISD/aug-cc-pVDZ} \\
\midrule
Atom & \multicolumn{1}{c}{$x$} & \multicolumn{1}{c}{$y$} & \multicolumn{1}{c}{$z$} \\
\midrule
C & 0.6026370736 & $-$0.4643907097 & 0.6583477049 \\
C & 0.6107315656 & $-$0.4625418977 & $-$2.1130618791 \\
O & 0.6098140206 & 1.8547083643 & $-$0.7220924201 \\
C & $-$1.7414923724 & $-$1.0330802257 & 2.1654074609 \\
H & 2.3921149876 & $-$0.8490307387 & 1.6050488609 \\
H & $-$1.1357319524 & $-$0.8634548347 & $-$3.1238551091 \\
H & 2.3489102876 & $-$0.8549518147 & $-$3.1384603491 \\
H & $-$3.4335192224 & $-$0.6015376227 & 1.0575924509 \\
H & $-$1.7806147124 & 0.1044763570 & 3.8947363309 \\
H & $-$1.7810716124 & $-$3.0335319957 & 2.7029653209 \\
\bottomrule
\end{tabular}
\end{minipage}%
\hfill
\begin{minipage}[t]{0.49\textwidth}\centering
\setlength{\tabcolsep}{4pt}
\begin{tabular}{lrrr}
\toprule
\multicolumn{4}{c}{CISD/cc-pVTZ} \\
\midrule
Atom & \multicolumn{1}{c}{$x$} & \multicolumn{1}{c}{$y$} & \multicolumn{1}{c}{$z$} \\
\midrule
C & 0.3719834149 & 0.0008026960 & 0.9243852147 \\
C & $-$1.8529186381 & 1.2283982496 & $-$0.1166444209 \\
O & $-$1.4554368596 & $-$1.3912959949 & $-$0.4271003967 \\
C & 2.9178515860 & 0.2409101861 & $-$0.2779092820 \\
H & 0.3734190264 & $-$0.3893813362 & 2.9240555374 \\
H & $-$1.6710672857 & 2.3377654050 & $-$1.8122950710 \\
H & $-$3.4025399289 & 1.7322825104 & 1.0994613745 \\
H & 2.7264226611 & 0.6305783469 & $-$2.2761564997 \\
H & 3.9818917040 & $-$1.4937018937 & $-$0.0577833400 \\
H & 3.9815960042 & 1.7589726527 & 0.5924981900 \\
\bottomrule
\end{tabular}
\end{minipage}%
\par\bigskip
\subsection{($S$)-3-Chloro-1-butyne}
\noindent
\begin{minipage}[t]{0.49\textwidth}\centering
\setlength{\tabcolsep}{4pt}
\begin{tabular}{lrrr}
\toprule
\multicolumn{4}{c}{MP2/cc-pVDZ} \\
\midrule
Atom & \multicolumn{1}{c}{$x$} & \multicolumn{1}{c}{$y$} & \multicolumn{1}{c}{$z$} \\
\midrule
C & $-$0.0642881940 & $-$0.8179496367 & $-$0.8405844105 \\
C & $-$2.4852963314 & 0.3732721442 & $-$0.2148565814 \\
Cl & 2.5216430575 & 1.2065972292 & 0.1377809260 \\
C & 0.2280940290 & $-$3.4176600336 & 0.3820835304 \\
H & 0.1182000274 & $-$0.9646106471 & $-$2.9100359010 \\
C & $-$4.5746370643 & 1.2724668417 & 0.2828590012 \\
H & $-$6.3746870599 & 2.1001871391 & 0.7388185172 \\
H & $-$1.3129681502 & $-$4.6612531432 & $-$0.2661321806 \\
H & 2.0562994116 & $-$4.2575527884 & $-$0.1465062866 \\
H & 0.1298464736 & $-$3.2453441574 & 2.4528218157 \\
\bottomrule
\end{tabular}
\end{minipage}%
\hfill
\begin{minipage}[t]{0.49\textwidth}\centering
\setlength{\tabcolsep}{4pt}
\begin{tabular}{lrrr}
\toprule
\multicolumn{4}{c}{MP2/aug-cc-pVDZ} \\
\midrule
Atom & \multicolumn{1}{c}{$x$} & \multicolumn{1}{c}{$y$} & \multicolumn{1}{c}{$z$} \\
\midrule
C & $-$0.0712379945 & $-$0.8381392299 & $-$0.8551028190 \\
C & $-$2.4858155119 & 0.3691914974 & $-$0.2418867645 \\
Cl & 2.5191068046 & 1.2119908216 & 0.1395851553 \\
C & 0.2255161252 & $-$3.4238706183 & 0.3961180359 \\
H & 0.1404784338 & $-$0.9831315919 & $-$2.9178514708 \\
C & $-$4.5607266827 & 1.2863091192 & 0.2997210186 \\
H & $-$6.3491974804 & 2.1282616776 & 0.7603415737 \\
H & $-$1.3000639317 & $-$4.6797004964 & $-$0.2588160303 \\
H & 2.0625768966 & $-$4.2510060799 & $-$0.1116048315 \\
H & 0.1048960304 & $-$3.2320281130 & 2.4611271126 \\
\bottomrule
\end{tabular}
\end{minipage}%
\par\bigskip
\noindent
\begin{minipage}[t]{0.49\textwidth}\centering
\setlength{\tabcolsep}{4pt}
\begin{tabular}{lrrr}
\toprule
\multicolumn{4}{c}{MP2/cc-pVTZ} \\
\midrule
Atom & \multicolumn{1}{c}{$x$} & \multicolumn{1}{c}{$y$} & \multicolumn{1}{c}{$z$} \\
\midrule
C & $-$0.0675291154 & $-$0.8233562853 & $-$0.8454151773 \\
C & $-$2.4677069510 & 0.3607591860 & $-$0.2303891600 \\
Cl & 2.4923465076 & 1.2015030262 & 0.1377248651 \\
C & 0.2221593818 & $-$3.3998726837 & 0.3885298475 \\
H & 0.1277548049 & $-$0.9729625849 & $-$2.8863498398 \\
C & $-$4.5051419278 & 1.2802364715 & 0.2921719464 \\
H & $-$6.2742162421 & 2.1064410825 & 0.7497958162 \\
H & $-$1.2946364302 & $-$4.6352363786 & $-$0.2504164126 \\
H & 2.0331050920 & $-$4.2265151556 & $-$0.1188164530 \\
H & 0.1135474734 & $-$3.2144821332 & 2.4315180497 \\
\bottomrule
\end{tabular}
\end{minipage}%
\hfill
\begin{minipage}[t]{0.49\textwidth}\centering
\setlength{\tabcolsep}{4pt}
\begin{tabular}{lrrr}
\toprule
\multicolumn{4}{c}{CISD/cc-pVDZ} \\
\midrule
Atom & \multicolumn{1}{c}{$x$} & \multicolumn{1}{c}{$y$} & \multicolumn{1}{c}{$z$} \\
\midrule
C & $-$0.0067371056 & 0.8152843279 & 0.8381423770 \\
C & $-$2.5097602782 & $-$0.2082617214 & 0.2225044784 \\
Cl & 2.4235948118 & $-$1.3686432214 & $-$0.1651734186 \\
C & 0.4494911248 & 3.3967386786 & $-$0.3539578486 \\
H & 0.1870536438 & 0.9242228756 & 2.8929288614 \\
C & $-$4.6085663382 & $-$0.9686454214 & $-$0.2491356386 \\
H & $-$6.4545521082 & $-$1.6698723114 & $-$0.6837025686 \\
H & $-$0.9916798522 & 4.7260773386 & 0.3171817234 \\
H & 2.3214671218 & 4.1026830186 & 0.1719103794 \\
H & 0.3303055228 & 3.2665702786 & $-$2.4152488986 \\
\bottomrule
\end{tabular}
\end{minipage}%
\par\bigskip
\noindent
\begin{minipage}[t]{0.49\textwidth}\centering
\setlength{\tabcolsep}{4pt}
\begin{tabular}{lrrr}
\toprule
\multicolumn{4}{c}{CISD/aug-cc-pVDZ} \\
\midrule
Atom & \multicolumn{1}{c}{$x$} & \multicolumn{1}{c}{$y$} & \multicolumn{1}{c}{$z$} \\
\midrule
C & $-$0.0085155232 & 0.8242227170 & 0.8474723490 \\
C & $-$2.5095561595 & $-$0.2070728280 & 0.2385722030 \\
Cl & 2.4202429305 & $-$1.3729240530 & $-$0.1683847980 \\
C & 0.4511603445 & 3.3995692070 & $-$0.3566565880 \\
H & 0.2029999125 & 0.9285368480 & 2.8957422970 \\
C & $-$4.6020186595 & $-$0.9686475540 & $-$0.2608297220 \\
H & $-$6.4399996995 & $-$1.6737675630 & $-$0.6999350760 \\
H & $-$0.9779438435 & 4.7332231670 & 0.3212895400 \\
H & 2.3265193605 & 4.0932704270 & 0.1589306080 \\
H & 0.3182276845 & 3.2626907670 & $-$2.4125646830 \\
\bottomrule
\end{tabular}
\end{minipage}%
\hfill
\begin{minipage}[t]{0.49\textwidth}\centering
\setlength{\tabcolsep}{4pt}
\begin{tabular}{lrrr}
\toprule
\multicolumn{4}{c}{CISD/cc-pVTZ} \\
\midrule
Atom & \multicolumn{1}{c}{$x$} & \multicolumn{1}{c}{$y$} & \multicolumn{1}{c}{$z$} \\
\midrule
C & $-$0.0058432642 & 0.8124742037 & 0.8364917794 \\
C & $-$2.4927101580 & $-$0.1961727885 & 0.2314260688 \\
Cl & 2.3949371620 & $-$1.3665891485 & $-$0.1663567372 \\
C & 0.4500501560 & 3.3768735615 & $-$0.3527592772 \\
H & 0.1943249050 & 0.9222247105 & 2.8626161328 \\
C & $-$4.5513673980 & $-$0.9582666285 & $-$0.2524036472 \\
H & $-$6.3689432580 & $-$1.6510666285 & $-$0.6849389772 \\
H & $-$0.9673148910 & 4.6937930115 & 0.3113575748 \\
H & 2.3015451420 & 4.0686276415 & 0.1598900458 \\
H & 0.3259568400 & 3.2473024415 & $-$2.3867312972 \\
\bottomrule
\end{tabular}
\end{minipage}%
\par\bigskip

\clearpage
\section{($P$)-H$_2$O$_2$}
\label{si:h2o2}

\begin{table}[htbp]
\caption{MP2 VCD harmonic vibrational frequencies (cm$^{-1}$), LG, VG, and
mixed LG/VG dipole strengths (10$^{-40}$ esu$^{2}$ cm$^{2}$), and LG, VG,
and LG(OI) rotatory strengths (10$^{-44}$ esu$^{2}$ cm$^{2}$) with
respect to basis set for each of the six vibrational modes of ($P$)-\ce{H2O2}.
LG rotatory strengths were computed with the molecular center of mass as
the gauge origin.}
\label{tab:si_h2o2_mp2}
\scriptsize
    \renewcommand{\arraystretch}{0.6}
    \begin{tabular*}{\textwidth}{@{\extracolsep{\fill}}llrrrrrrrrr}
    \toprule
    & & & & \multicolumn{3}{c}{Dipole Strength} & \hspace{1pt} & \multicolumn{3}{c}{Rotatory Strength} \\
    \cmidrule{5-7} \cmidrule{9-11}
    Mode & Sym. & \multicolumn{1}{c}{Frequency} & & \multicolumn{1}{c}{LG} & \multicolumn{1}{c}{VG} & \multicolumn{1}{c}{Mixed} & & \multicolumn{1}{c}{LG} & \multicolumn{1}{c}{VG} & \multicolumn{1}{c}{LG(OI)} \\ \midrule
    \multicolumn{11}{c}{MP2/cc-pVDZ} \\
    \cmidrule{1-11}
    1 & A & 337.9 &  & 2271.987 & 1371.687 & 1765.348 &  & 143.485 & 111.489 & 143.485 \\
    2 & A & 920.4 &  & 5.586 & 1.273 & 2.667 &  & $-$2.398 & $-$1.145 & $-$2.398 \\
    3 & B & 1306.0 &  & 349.335 & 313.911 & 326.311 &  & 4.529 & 32.436 & 34.218 \\
    4 & A & 1442.3 &  & 0.300 & 17.318 & 2.280 &  & $-$7.812 & $-$59.331 & $-$7.812 \\
    5 & B & 3807.2 &  & 59.632 & 8.415 & 8.161 &  & $-$38.393 & 3.502 & 9.321 \\
    6 & A & 3809.7 &  & 14.354 & 0.699 & $-$3.167 &  & 24.984 & $-$5.513 & $-$24.984 \\
    \cmidrule{9-11}
    \multicolumn{3}{l}{RMS dev.\ from VG} &  &  &  &  &  & 34.50 & \multicolumn{1}{c}{---} & 26.13 \\
    \cmidrule{1-11}
    \multicolumn{11}{c}{MP2/cc-pVTZ} \\
    \cmidrule{1-11}
    1 & A & 372.0 &  & 1865.274 & 1483.674 & 1663.568 &  & 161.829 & 144.329 & 161.829 \\
    2 & A & 944.6 &  & 3.523 & 3.651 & 3.586 &  & $-$4.152 & $-$4.227 & $-$4.152 \\
    3 & B & 1328.0 &  & 352.214 & 229.022 & 283.773 &  & 9.322 & 14.365 & 17.815 \\
    4 & A & 1437.4 &  & 0.461 & 2.185 & 1.004 &  & $-$10.386 & $-$22.604 & $-$10.386 \\
    5 & B & 3819.3 &  & 62.436 & 39.967 & 49.544 &  & $-$21.852 & $-$15.644 & $-$19.553 \\
    6 & A & 3820.3 &  & 17.524 & 7.193 & 11.227 &  & 10.823 & 6.934 & 10.823 \\
    \cmidrule{9-11}
    \multicolumn{3}{l}{RMS dev.\ from VG} &  &  &  &  &  & 9.44 & \multicolumn{1}{c}{---} & 9.11 \\
    \cmidrule{1-11}
    \multicolumn{11}{c}{MP2/cc-pVQZ} \\
    \cmidrule{1-11}
    1 & A & 384.1 &  & 1758.568 & 1688.046 & 1722.946 &  & 176.596 & 173.019 & 176.596 \\
    2 & A & 937.6 &  & 2.930 & 4.614 & 3.677 &  & $-$4.108 & $-$5.155 & $-$4.108 \\
    3 & B & 1330.8 &  & 345.442 & 278.446 & 310.018 &  & 12.642 & 16.911 & 18.835 \\
    4 & A & 1435.3 &  & 0.726 & 1.246 & 0.951 &  & $-$13.718 & $-$17.966 & $-$13.718 \\
    5 & B & 3817.4 &  & 67.568 & 70.165 & 68.823 &  & $-$13.008 & $-$12.905 & $-$12.664 \\
    6 & A & 3817.6 &  & 19.476 & 18.141 & 18.796 &  & 2.245 & 2.167 & 2.245 \\
    \cmidrule{9-11}
    \multicolumn{3}{l}{RMS dev.\ from VG} &  &  &  &  &  & 2.89 & \multicolumn{1}{c}{---} & 2.44 \\
    \cmidrule{1-11}
    \multicolumn{11}{c}{MP2/aug-cc-pVDZ} \\
    \cmidrule{1-11}
    1 & A & 394.2 &  & 1669.904 & 900.926 & 1226.564 &  & 138.919 & 102.037 & 138.919 \\
    2 & A & 881.3 &  & 2.358 & 0.977 & 1.518 &  & $-$2.759 & $-$1.776 & $-$2.759 \\
    3 & B & 1309.9 &  & 344.942 & 116.645 & 200.380 &  & 15.264 & 14.063 & 24.184 \\
    4 & A & 1414.0 &  & 0.755 & 1.045 & 0.888 &  & $-$12.112 & $-$14.248 & $-$12.112 \\
    5 & B & 3768.8 &  & 59.412 & 31.670 & 41.782 &  & $-$18.349 & $-$10.222 & $-$14.001 \\
    6 & A & 3767.5 &  & 16.157 & 4.359 & 8.393 &  & 8.738 & 4.539 & 8.738 \\
    \cmidrule{9-11}
    \multicolumn{3}{l}{RMS dev.\ from VG} &  &  &  &  &  & 15.55 & \multicolumn{1}{c}{---} & 15.81 \\
    \cmidrule{1-11}
    \multicolumn{11}{c}{MP2/aug-cc-pVTZ} \\
    \cmidrule{1-11}
    1 & A & 385.2 &  & 1694.634 & 1319.303 & 1495.237 &  & 160.526 & 141.638 & 160.526 \\
    2 & A & 926.5 &  & 1.968 & 2.245 & 2.102 &  & $-$3.225 & $-$3.445 & $-$3.225 \\
    3 & B & 1319.6 &  & 340.029 & 185.169 & 250.917 &  & 13.674 & 11.253 & 15.249 \\
    4 & A & 1422.4 &  & 0.776 & 0.412 & 0.565 &  & $-$13.440 & $-$9.795 & $-$13.440 \\
    5 & B & 3787.5 &  & 66.453 & 59.157 & 62.460 &  & $-$12.842 & $-$11.172 & $-$11.840 \\
    6 & A & 3786.8 &  & 18.374 & 14.097 & 16.094 &  & 3.135 & 2.746 & 3.135 \\
    \cmidrule{9-11}
    \multicolumn{3}{l}{RMS dev.\ from VG} &  &  &  &  &  & 7.95 & \multicolumn{1}{c}{---} & 8.03 \\
    \cmidrule{1-11}
    \multicolumn{11}{c}{MP2/aug-cc-pVQZ} \\
    \cmidrule{1-11}
    1 & A & 384.3 &  & 1717.354 & 1660.659 & 1688.769 &  & 175.240 & 172.324 & 175.240 \\
    2 & A & 932.7 &  & 2.081 & 3.411 & 2.664 &  & $-$3.518 & $-$4.505 & $-$3.518 \\
    3 & B & 1328.6 &  & 338.334 & 258.626 & 295.778 &  & 13.358 & 14.419 & 16.492 \\
    4 & A & 1431.8 &  & 0.742 & 0.450 & 0.578 &  & $-$13.957 & $-$10.874 & $-$13.957 \\
    5 & B & 3805.8 &  & 69.325 & 79.530 & 74.209 &  & $-$8.721 & $-$9.011 & $-$8.413 \\
    6 & A & 3805.1 &  & 19.098 & 21.290 & 20.164 &  & $-$1.006 & $-$1.062 & $-$1.006 \\
    \cmidrule{9-11}
    \multicolumn{3}{l}{RMS dev.\ from VG} &  &  &  &  &  & 1.83 & \multicolumn{1}{c}{---} & 1.99 \\
    \bottomrule
    \end{tabular*}

\end{table}

\begin{table}[htbp]
\caption{CISD VCD harmonic vibrational frequencies (cm$^{-1}$), LG, VG, and
mixed LG/VG dipole strengths (10$^{-40}$ esu$^{2}$ cm$^{2}$), and LG, VG,
and LG(OI) rotatory strengths (10$^{-44}$ esu$^{2}$ cm$^{2}$) with
respect to basis set for each of the six vibrational modes of ($P$)-\ce{H2O2}.
LG rotatory strengths were computed with the molecular center of mass as
the gauge origin.}
\label{tab:si_h2o2_cisd}
\scriptsize
    \renewcommand{\arraystretch}{0.6}
    \begin{tabular*}{\textwidth}{@{\extracolsep{\fill}}llrrrrrrrrr}
    \toprule
    & & & & \multicolumn{3}{c}{Dipole Strength} & \hspace{1pt} & \multicolumn{3}{c}{Rotatory Strength} \\
    \cmidrule{5-7} \cmidrule{9-11}
    Mode & Sym. & \multicolumn{1}{c}{Frequency} & & \multicolumn{1}{c}{LG} & \multicolumn{1}{c}{VG} & \multicolumn{1}{c}{Mixed} & & \multicolumn{1}{c}{LG} & \multicolumn{1}{c}{VG} & \multicolumn{1}{c}{LG(OI)} \\ \midrule
    \multicolumn{11}{c}{CISD/cc-pVDZ} \\
    \cmidrule{1-11}
    1 & A & 363.0 &  & 2101.640 & 1118.808 & 1533.405 &  & 155.762 & 113.647 & 155.762 \\
    2 & A & 994.2 &  & 5.690 & 0.084 & 0.690 &  & $-$2.029 & $-$0.246 & $-$2.029 \\
    3 & B & 1369.5 &  & 296.761 & 301.045 & 295.550 &  & 5.147 & 29.781 & 29.569 \\
    4 & A & 1503.3 &  & 0.472 & 16.972 & 2.831 &  & $-$9.667 & $-$57.962 & $-$9.667 \\
    5 & B & 3910.9 &  & 64.732 & 9.969 & $-$3.442 &  & $-$60.795 & 18.515 & 47.180 \\
    6 & A & 3916.0 &  & 16.935 & 3.413 & $-$7.603 &  & 41.498 & $-$18.630 & $-$41.498 \\
    \cmidrule{9-11}
    \multicolumn{3}{l}{RMS dev.\ from VG} &  &  &  &  &  & 49.37 & \multicolumn{1}{c}{---} & 30.15 \\
    \cmidrule{1-11}
    \multicolumn{11}{c}{CISD/cc-pVTZ} \\
    \cmidrule{1-11}
    1 & A & 395.7 &  & 1800.031 & 1318.702 & 1540.683 &  & 177.081 & 151.567 & 177.081 \\
    2 & A & 1048.5 &  & 3.472 & 1.681 & 2.416 &  & $-$3.699 & $-$2.574 & $-$3.699 \\
    3 & B & 1401.0 &  & 305.019 & 215.741 & 256.508 &  & 11.433 & 11.484 & 13.654 \\
    4 & A & 1512.6 &  & 0.687 & 2.200 & 1.230 &  & $-$12.577 & $-$22.506 & $-$12.577 \\
    5 & B & 3961.1 &  & 68.802 & 29.209 & 44.230 &  & $-$38.424 & $-$21.849 & $-$33.534 \\
    6 & A & 3964.9 &  & 20.089 & 4.754 & 9.773 &  & 22.605 & 10.997 & 22.605 \\
    \cmidrule{9-11}
    \multicolumn{3}{l}{RMS dev.\ from VG} &  &  &  &  &  & 13.91 & \multicolumn{1}{c}{---} & 13.08 \\
    \cmidrule{1-11}
    \multicolumn{11}{c}{CISD/cc-pVQZ} \\
    \cmidrule{1-11}
    1 & A & 406.2 &  & 1732.533 & 1548.673 & 1638.025 &  & 192.450 & 181.952 & 192.450 \\
    2 & A & 1054.9 &  & 2.756 & 2.496 & 2.623 &  & $-$3.634 & $-$3.459 & $-$3.634 \\
    3 & B & 1409.5 &  & 302.416 & 262.156 & 281.565 &  & 14.754 & 14.456 & 15.526 \\
    4 & A & 1517.6 &  & 0.868 & 1.329 & 1.074 &  & $-$14.983 & $-$18.538 & $-$14.983 \\
    5 & B & 3971.2 &  & 74.673 & 58.291 & 65.958 &  & $-$28.159 & $-$24.466 & $-$27.691 \\
    6 & A & 3974.4 &  & 21.937 & 15.190 & 18.254 &  & 12.386 & 10.307 & 12.386 \\
    \cmidrule{9-11}
    \multicolumn{3}{l}{RMS dev.\ from VG} &  &  &  &  &  & 4.85 & \multicolumn{1}{c}{---} & 4.81 \\
    \cmidrule{1-11}
    \multicolumn{11}{c}{CISD/aug-cc-pVDZ} \\
    \cmidrule{1-11}
    1 & A & 415.1 &  & 1643.822 & 789.160 & 1138.964 &  & 151.133 & 104.716 & 151.133 \\
    2 & A & 985.7 &  & 2.409 & 0.177 & 0.652 &  & $-$2.646 & $-$0.717 & $-$2.646 \\
    3 & B & 1383.7 &  & 294.881 & 107.905 & 178.374 &  & 15.862 & 10.396 & 17.186 \\
    4 & A & 1491.6 &  & 0.848 & 1.015 & 0.928 &  & $-$12.524 & $-$13.701 & $-$12.524 \\
    5 & B & 3909.2 &  & 64.527 & 18.409 & 32.857 &  & $-$37.519 & $-$14.895 & $-$27.886 \\
    6 & A & 3912.6 &  & 17.747 & 1.878 & 5.772 &  & 22.155 & 7.206 & 22.155 \\
    \cmidrule{9-11}
    \multicolumn{3}{l}{RMS dev.\ from VG} &  &  &  &  &  & 22.08 & \multicolumn{1}{c}{---} & 20.81 \\
    \cmidrule{1-11}
    \multicolumn{11}{c}{CISD/aug-cc-pVTZ} \\
    \cmidrule{1-11}
    1 & A & 406.1 &  & 1687.231 & 1200.762 & 1423.363 &  & 175.141 & 147.750 & 175.141 \\
    2 & A & 1042.2 &  & 2.087 & 1.062 & 1.489 &  & $-$3.070 & $-$2.191 & $-$3.070 \\
    3 & B & 1398.7 &  & 296.755 & 173.571 & 226.912 &  & 15.137 & 8.716 & 11.397 \\
    4 & A & 1505.8 &  & 0.806 & 0.412 & 0.576 &  & $-$13.607 & $-$9.725 & $-$13.607 \\
    5 & B & 3942.4 &  & 73.058 & 46.216 & 57.982 &  & $-$28.707 & $-$21.721 & $-$27.310 \\
    6 & A & 3945.6 &  & 20.544 & 10.977 & 15.017 &  & 13.793 & 10.082 & 13.793 \\
    \cmidrule{9-11}
    \multicolumn{3}{l}{RMS dev.\ from VG} &  &  &  &  &  & 12.04 & \multicolumn{1}{c}{---} & 11.68 \\
    \cmidrule{1-11}
    \multicolumn{11}{c}{CISD/aug-cc-pVQZ} \\
    \cmidrule{1-11}
    1 & A & 405.6 &  & 1710.590 & 1529.391 & 1617.455 &  & 191.010 & 180.610 & 191.010 \\
    2 & A & 1053.9 &  & 2.144 & 1.843 & 1.988 &  & $-$3.293 & $-$3.054 & $-$3.293 \\
    3 & B & 1409.3 &  & 297.412 & 243.131 & 268.897 &  & 15.261 & 12.344 & 13.652 \\
    4 & A & 1516.8 &  & 0.771 & 0.504 & 0.624 &  & $-$14.229 & $-$11.508 & $-$14.229 \\
    5 & B & 3965.2 &  & 76.277 & 66.969 & 71.469 &  & $-$23.430 & $-$21.807 & $-$23.273 \\
    6 & A & 3968.0 &  & 21.456 & 18.285 & 19.807 &  & 8.615 & 7.953 & 8.615 \\
    \cmidrule{9-11}
    \multicolumn{3}{l}{RMS dev.\ from VG} &  &  &  &  &  & 4.60 & \multicolumn{1}{c}{---} & 4.47 \\
    \bottomrule
    \end{tabular*}

\end{table}

\clearpage
\section{($S$)-Methyloxirane}
\label{si:smo}

\begin{figure}[htbp]
\centering
\includegraphics[width=\textwidth]{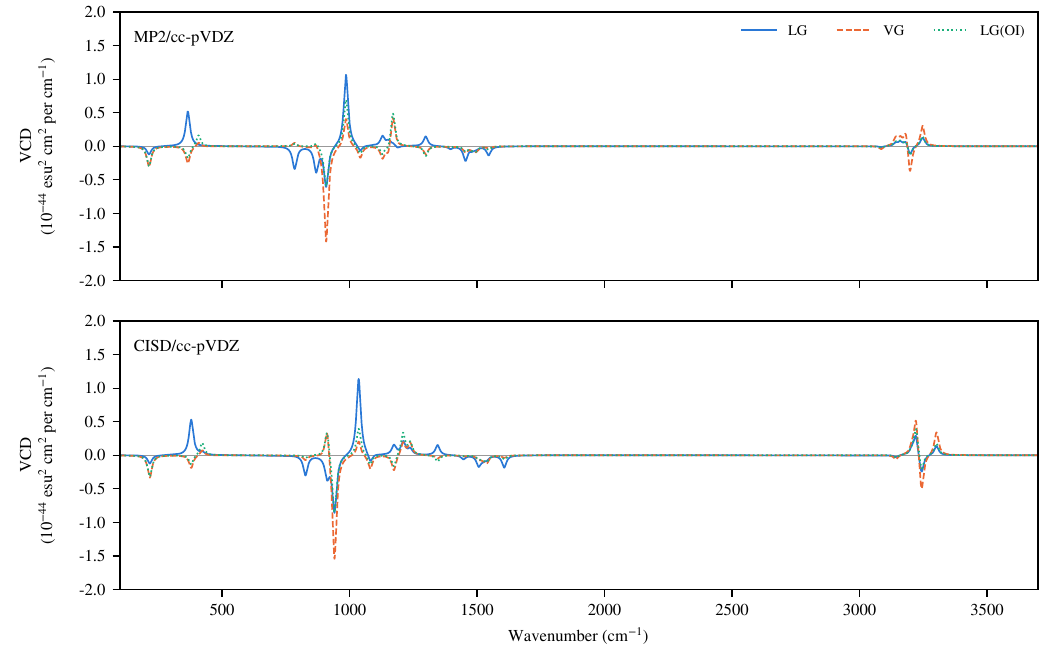}
\caption{MP2 (top) and CISD (bottom) VCD spectra of ($S$)-methyloxirane computed with the cc-pVDZ basis set,
showing the LG (blue, solid), VG (orange, dashed), and LG(OI) (green,
dotted) rotatory strengths. Each band is broadened with an area-normalized
Lorentzian function with a half-width at half-maximum of 10~cm$^{-1}$.}
\label{fig:si_smo_pvdz}
\end{figure}

\begin{figure}[htbp]
\centering
\includegraphics[width=\textwidth]{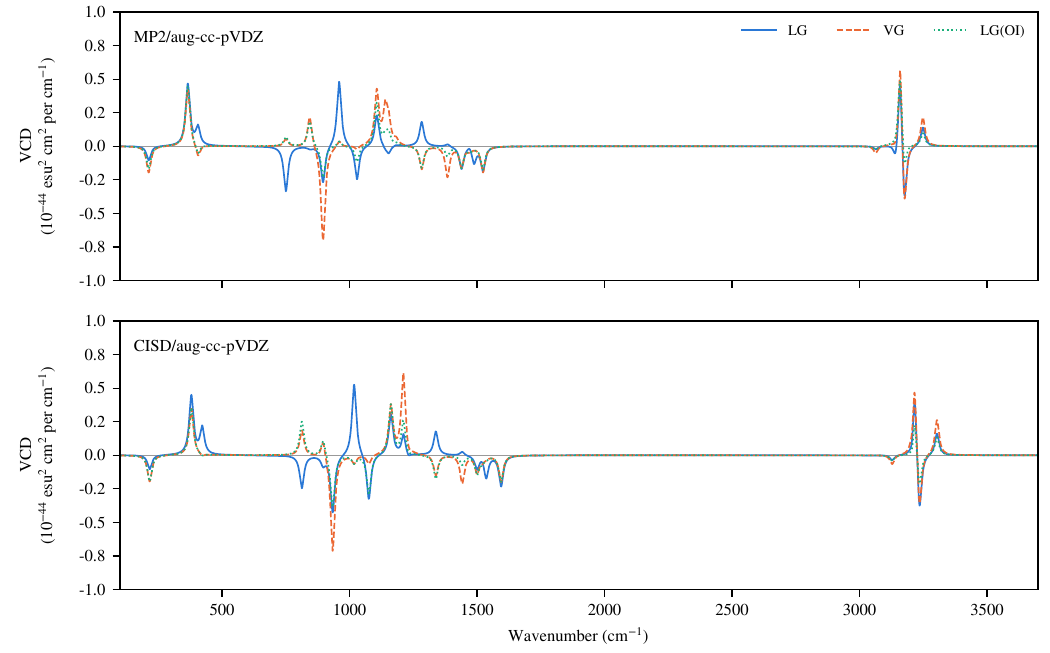}
\caption{MP2 (top) and CISD (bottom) VCD spectra of ($S$)-methyloxirane computed with the aug-cc-pVDZ basis set,
showing the LG (blue, solid), VG (orange, dashed), and LG(OI) (green,
dotted) rotatory strengths. Each band is broadened with an area-normalized
Lorentzian function with a half-width at half-maximum of 10~cm$^{-1}$.}
\label{fig:si_smo_apvdz}
\end{figure}

\begin{table}[htbp]
\caption{MP2 VCD harmonic vibrational frequencies (cm$^{-1}$), LG, VG, and
mixed LG/VG dipole strengths (10$^{-40}$ esu$^{2}$ cm$^{2}$), LG, VG, and
LG(OI) rotatory strengths (10$^{-44}$ esu$^{2}$ cm$^{2}$), and the cosine of
the angle between the LG and VG APT vectors, $\cos\theta$, with respect to
basis set for each of the 24 vibrational modes of ($S$)-methyloxirane. LG rotatory
strengths were computed with the molecular center of mass as the gauge
origin. Modes are numbered by ascending frequency at the CISD/cc-pVTZ
level.}
\label{tab:si_smo_mp2}
\scriptsize
    \renewcommand{\arraystretch}{0.6}
    \begin{tabular*}{\textwidth}{@{\extracolsep{\fill}}lrcrrrcrrrr}
    \toprule
    & & & \multicolumn{3}{c}{Dipole Strength} & \hspace{1pt} & \multicolumn{3}{c}{Rotatory Strength} & \\
    \cmidrule{4-6} \cmidrule{8-10}
    Mode & \multicolumn{1}{c}{$\nu$} &  & \multicolumn{1}{c}{LG} & \multicolumn{1}{c}{VG} & \multicolumn{1}{c}{Mixed} &  & \multicolumn{1}{c}{LG} & \multicolumn{1}{c}{VG} & \multicolumn{1}{c}{LG(OI)} & \multicolumn{1}{c}{$\cos\theta$} \\ \midrule
    \multicolumn{11}{c}{MP2/cc-pVDZ} \\
    \cmidrule{1-11}
    1 & 213.6 &  & 10.402 & 11.216 & 3.289 &  & $-$3.864 & $-$9.628 & $-$9.272 & 0.30 \\
    2 & 366.0 &  & 39.850 & 76.685 & $-$21.885 &  & 16.301 & $-$8.104 & $-$5.842 & $-$0.40 \\
    3 & 407.9 &  & 41.006 & 6.381 & $-$13.033 &  & $-$0.137 & 2.141 & 5.428 & $-$0.81 \\
    4 & 784.8 &  & 26.156 & 57.365 & $-$19.701 &  & $-$10.457 & 1.812 & 1.224 & $-$0.51 \\
    5 & 869.3 &  & 152.143 & 293.075 & 203.283 &  & $-$11.311 & 3.013 & 2.171 & 0.96 \\
    6 & 908.4 &  & 9.651 & 53.317 & 18.186 &  & $-$18.831 & $-$44.971 & $-$19.133 & 0.80 \\
    7 & 986.2 &  & 39.360 & 14.857 & 6.974 &  & 33.877 & 13.617 & 22.164 & 0.29 \\
    8 & 1042.1 &  & 34.868 & 69.327 & 49.132 &  & $-$2.932 & $-$5.532 & $-$3.924 & 1.00 \\
    9 & 1129.3 &  & 21.501 & 41.497 & 23.037 &  & 4.539 & $-$5.784 & $-$4.163 & 0.77 \\
    10 & 1170.3\smash{$^{\rm a}$} &  & 11.182 & 9.291 & 2.844 &  & 1.002 & 14.854 & 16.295 & 0.28 \\
    11 & 1152.3\smash{$^{\rm a}$} &  & 1.871 & 5.527 & 0.932 &  & 2.147 & $-$5.031 & $-$2.928 & 0.29 \\
    12 & 1188.5 &  & 2.144 & 2.804 & 2.030 &  & $-$1.152 & $-$0.525 & $-$0.459 & 0.83 \\
    13 & 1298.6 &  & 9.615 & 6.565 & $-$4.308 &  & 4.631 & $-$3.834 & $-$4.640 & $-$0.54 \\
    14 & 1395.3 &  & 10.559 & 22.487 & 14.019 &  & $-$1.226 & $-$0.438 & $-$0.300 & 0.91 \\
    15 & 1454.8 &  & 45.258 & 67.708 & 51.469 &  & $-$6.625 & $-$2.830 & $-$2.314 & 0.93 \\
    16 & 1480.5 &  & 13.138 & 8.396 & 10.328 &  & $-$1.405 & $-$1.131 & $-$1.414 & 0.98 \\
    17 & 1496.1 &  & 14.939 & 20.220 & 16.548 &  & $-$1.001 & $-$2.324 & $-$1.998 & 0.95 \\
    18 & 1545.6 &  & 16.622 & 6.465 & 9.615 &  & $-$4.158 & $-$0.943 & $-$1.512 & 0.93 \\
    19 & 3085.1 &  & 21.228 & 83.627 & 42.002 &  & $-$0.503 & $-$1.493 & $-$0.752 & 1.00 \\
    20 & 3142.2 &  & 27.474 & 111.448 & 55.326 &  & 1.381 & 3.672 & 1.823 & 1.00 \\
    21 & 3182.0\smash{$^{\rm a}$} &  & 19.605 & 172.360 & 58.098 &  & 2.762 & 8.738 & 2.947 & 1.00 \\
    22 & 3159.7\smash{$^{\rm ab}$} &  & 23.162 & 61.112 & 37.599 &  & 1.912 & 3.355 & 2.065 & 1.00 \\
    23 & 3197.4\smash{$^{\rm b}$} &  & 24.841 & 199.200 & 70.340 &  & $-$4.639 & $-$14.741 & $-$5.206 & 1.00 \\
    24 & 3247.1 &  & 33.452 & 156.187 & 72.227 &  & 4.054 & 9.862 & 4.564 & 1.00 \\
    \cmidrule{8-10}
    \multicolumn{3}{l}{RMS dev.\ from VG} &  &  &  &  & 10.64 & \multicolumn{1}{c}{---} & 6.22 &  \\
    \multicolumn{3}{l}{$N_{\rm sign}$ LG/VG} &  &  &  &  & \multicolumn{3}{c}{7} &  \\
    \cmidrule{1-11}
    \multicolumn{11}{c}{MP2/aug-cc-pVDZ} \\
    \cmidrule{1-11}
    1 & 212.3 &  & 7.931 & 10.625 & 7.603 &  & $-$3.365 & $-$6.186 & $-$5.344 & 0.83 \\
    2 & 365.9 &  & 47.168 & 41.660 & 6.854 &  & 14.451 & 13.458 & 14.320 & 0.15 \\
    3 & 405.5 &  & 39.716 & 55.409 & 32.060 &  & 4.247 & $-$2.930 & $-$2.480 & 0.68 \\
    4 & 750.5 &  & 37.226 & 24.643 & 26.794 &  & $-$10.545 & 1.759 & 2.162 & 0.88 \\
    5 & 843.8 &  & 218.940 & 343.676 & 270.985 &  & $-$0.482 & 7.507 & 5.991 & 0.99 \\
    6 & 896.3 &  & 6.175 & 54.038 & 14.366 &  & $-$8.697 & $-$22.378 & $-$7.564 & 0.79 \\
    7 & 959.2 &  & 37.347 & 58.686 & 32.560 &  & 15.455 & 1.562 & 1.246 & 0.70 \\
    8 & 1029.4 &  & 29.103 & 1.639 & 5.884 &  & $-$8.071 & $-$0.907 & $-$3.821 & 0.85 \\
    9 & 1106.9 &  & 19.592 & 31.390 & 23.721 &  & 7.396 & 12.640 & 9.986 & 0.96 \\
    10 & 1153.3\smash{$^{\rm ab}$} &  & 10.004 & 33.737 & 16.423 &  & $-$1.768 & 5.047 & 2.748 & 0.89 \\
    11 & 1140.3\smash{$^{\rm a}$} &  & 3.367 & 53.823 & $-$8.298 &  & $-$0.689 & 7.982 & 1.996 & $-$0.62 \\
    12 & 1183.1\smash{$^{\rm b}$} &  & 2.355 & 15.245 & 3.552 &  & 0.327 & 1.216 & 0.478 & 0.59 \\
    13 & 1283.1 &  & 9.822 & 11.058 & $-$3.364 &  & 5.726 & $-$5.413 & $-$5.102 & $-$0.32 \\
    14 & 1384.0 &  & 8.794 & 92.764 & 0.198 &  & 0.634 & $-$7.065 & $-$2.175 & 0.01 \\
    15 & 1439.2 &  & 46.303 & 43.430 & 41.938 &  & $-$5.236 & $-$4.731 & $-$4.885 & 0.94 \\
    16 & 1471.4 &  & 13.008 & 11.354 & $-$8.281 &  & 0.969 & $-$0.872 & $-$0.933 & $-$0.68 \\
    17 & 1488.2 &  & 16.835 & 6.369 & $-$1.246 &  & $-$3.833 & $-$0.111 & $-$0.181 & $-$0.12 \\
    18 & 1523.4 &  & 16.367 & 20.508 & 13.795 &  & $-$5.605 & $-$6.001 & $-$5.361 & 0.75 \\
    19 & 3063.5 &  & 20.672 & 65.271 & 36.702 &  & $-$0.724 & $-$1.656 & $-$0.932 & 1.00 \\
    20 & 3141.4 &  & 26.891 & 77.490 & 45.065 &  & $-$4.045 & $-$1.554 & $-$0.915 & 0.99 \\
    21 & 3154.3 &  & 17.353 & 115.043 & 44.641 &  & $-$6.290 & $-$9.767 & $-$3.793 & 1.00 \\
    22 & 3158.5 &  & 9.719 & 21.541 & 14.024 &  & 24.492 & 30.762 & 20.663 & 0.97 \\
    23 & 3175.9 &  & 38.734 & 189.695 & 85.516 &  & $-$16.051 & $-$17.875 & $-$8.077 & 1.00 \\
    24 & 3248.5 &  & 21.723 & 114.107 & 49.705 &  & 4.480 & 6.845 & 2.987 & 1.00 \\
    \cmidrule{8-10}
    \multicolumn{3}{l}{RMS dev.\ from VG} &  &  &  &  & 6.85 & \multicolumn{1}{c}{---} & 4.81 &  \\
    \multicolumn{3}{l}{$N_{\rm sign}$ LG/VG} &  &  &  &  & \multicolumn{3}{c}{7} &  \\
    \cmidrule{1-11}
    \multicolumn{11}{c}{MP2/cc-pVTZ} \\
    \cmidrule{1-11}
    1 & 218.4 &  & 8.663 & 9.646 & 8.695 &  & $-$3.853 & $-$6.177 & $-$5.854 & 0.95 \\
    2 & 368.4 &  & 37.971 & 18.897 & 21.217 &  & 13.104 & 11.350 & 16.088 & 0.79 \\
    3 & 408.4 &  & 44.572 & 25.026 & 31.315 &  & 3.909 & 1.447 & 1.931 & 0.94 \\
    4 & 784.8 &  & 29.218 & 16.675 & 21.452 &  & $-$9.771 & $-$4.722 & $-$6.251 & 0.97 \\
    5 & 867.4 &  & 172.558 & 310.598 & 229.505 &  & $-$4.274 & 2.937 & 2.189 & 0.99 \\
    6 & 913.7 &  & 7.918 & 31.651 & 13.903 &  & $-$15.822 & $-$33.107 & $-$16.559 & 0.88 \\
    7 & 984.7 &  & 45.350 & 59.022 & 44.867 &  & 25.072 & 20.073 & 17.595 & 0.87 \\
    8 & 1051.5 &  & 34.368 & 24.231 & 28.828 &  & $-$7.853 & $-$5.771 & $-$6.873 & 1.00 \\
    9 & 1135.3 &  & 18.995 & 27.429 & 22.821 &  & 7.143 & 8.636 & 7.187 & 1.00 \\
    10 & 1175.0\smash{$^{\rm a}$} &  & 11.872 & 19.079 & 13.660 &  & 3.456 & 12.464 & 9.833 & 0.91 \\
    11 & 1161.0\smash{$^{\rm a}$} &  & 1.702 & 11.023 & $-$2.542 &  & $-$1.627 & 1.653 & 0.650 & $-$0.59 \\
    12 & 1196.9 &  & 1.268 & 5.268 & 1.918 &  & $-$0.479 & $-$2.328 & $-$1.142 & 0.74 \\
    13 & 1304.1 &  & 11.786 & 3.778 & 4.669 &  & 6.645 & 1.564 & 2.762 & 0.70 \\
    14 & 1405.3 &  & 11.299 & 24.331 & 6.305 &  & $-$0.547 & $-$2.307 & $-$1.572 & 0.38 \\
    15 & 1457.6 &  & 52.500 & 51.504 & 51.337 &  & $-$9.565 & $-$7.929 & $-$8.006 & 0.99 \\
    16 & 1498.3 &  & 13.572 & 2.213 & 3.157 &  & $-$0.943 & $-$1.017 & $-$2.519 & 0.58 \\
    17 & 1513.7 &  & 16.053 & 5.790 & 8.032 &  & $-$1.182 & $-$1.343 & $-$2.236 & 0.83 \\
    18 & 1546.4 &  & 11.757 & 14.237 & 11.248 &  & $-$4.371 & $-$5.665 & $-$5.148 & 0.87 \\
    19 & 3077.5 &  & 17.287 & 49.915 & 29.342 &  & $-$0.519 & $-$1.231 & $-$0.724 & 1.00 \\
    20 & 3144.9 &  & 26.181 & 70.018 & 42.725 &  & $-$1.564 & $-$0.656 & $-$0.401 & 1.00 \\
    21 & 3164.4\smash{$^{\rm a}$} &  & 18.411 & 91.159 & 40.921 &  & 9.151 & 16.475 & 7.404 & 1.00 \\
    22 & 3162.1\smash{$^{\rm a}$} &  & 8.739 & 13.361 & 10.621 &  & 2.522 & 0.202 & 0.164 & 0.98 \\
    23 & 3182.1 &  & 32.337 & 134.343 & 65.880 &  & $-$9.621 & $-$13.678 & $-$6.711 & 1.00 \\
    24 & 3246.0 &  & 23.227 & 83.447 & 43.992 &  & 3.684 & 6.067 & 3.201 & 1.00 \\
    \cmidrule{8-10}
    \multicolumn{3}{l}{RMS dev.\ from VG} &  &  &  &  & 5.14 & \multicolumn{1}{c}{---} & 4.39 &  \\
    \multicolumn{3}{l}{$N_{\rm sign}$ LG/VG} &  &  &  &  & \multicolumn{3}{c}{2} &  \\
    \bottomrule
    \end{tabular*}
    \par\vspace{2pt}\noindent\parbox{\textwidth}{\raggedright
    $^{\rm a}$Frequency order differs from the reference level (CISD/cc-pVTZ); the row follows the matched normal mode.
    $^{\rm b}$Normal-mode overlap with the reference mode $<0.80$.
    RMS: root-mean-square difference of the rotatory strengths over all modes. $N_{\rm sign}$: number of modes whose LG and VG rotatory strengths differ in sign, counted for modes with $\max(|R_1|,|R_2|)\geq1\times10^{-44}$ esu$^2$ cm$^2$.}

\end{table}

\begin{table}[htbp]
\caption{CISD VCD harmonic vibrational frequencies (cm$^{-1}$), LG, VG, and
mixed LG/VG dipole strengths (10$^{-40}$ esu$^{2}$ cm$^{2}$), LG, VG, and
LG(OI) rotatory strengths (10$^{-44}$ esu$^{2}$ cm$^{2}$), and the cosine of
the angle between the LG and VG APT vectors, $\cos\theta$, with respect to
basis set for each of the 24 vibrational modes of ($S$)-methyloxirane. LG rotatory
strengths were computed with the molecular center of mass as the gauge
origin. Modes are numbered by ascending frequency at the CISD/cc-pVTZ
level.}
\label{tab:si_smo_cisd}
\scriptsize
    \renewcommand{\arraystretch}{0.6}
    \begin{tabular*}{\textwidth}{@{\extracolsep{\fill}}lrcrrrcrrrr}
    \toprule
    & & & \multicolumn{3}{c}{Dipole Strength} & \hspace{1pt} & \multicolumn{3}{c}{Rotatory Strength} & \\
    \cmidrule{4-6} \cmidrule{8-10}
    Mode & \multicolumn{1}{c}{$\nu$} &  & \multicolumn{1}{c}{LG} & \multicolumn{1}{c}{VG} & \multicolumn{1}{c}{Mixed} &  & \multicolumn{1}{c}{LG} & \multicolumn{1}{c}{VG} & \multicolumn{1}{c}{LG(OI)} & \multicolumn{1}{c}{$\cos\theta$} \\ \midrule
    \multicolumn{11}{c}{CISD/cc-pVDZ} \\
    \cmidrule{1-11}
    1 & 216.8 &  & 11.110 & 12.093 & 2.807 &  & $-$3.925 & $-$10.429 & $-$9.996 & 0.24 \\
    2 & 379.2 &  & 38.155 & 67.879 & $-$18.051 &  & 16.609 & $-$6.033 & $-$4.523 & $-$0.35 \\
    3 & 422.2 &  & 47.388 & 9.195 & $-$13.920 &  & 1.161 & 2.660 & 6.038 & $-$0.67 \\
    4 & 827.3 &  & 25.845 & 73.818 & 2.076 &  & $-$9.232 & $-$2.081 & $-$1.231 & 0.05 \\
    5 & 912.1 &  & 161.695 & 219.544 & 174.017 &  & $-$9.169 & 15.551 & 13.346 & 0.92 \\
    6 & 941.0 &  & 17.719 & 55.731 & 24.489 &  & $-$26.325 & $-$50.122 & $-$28.262 & 0.78 \\
    7 & 1035.9 &  & 50.718 & 15.992 & 8.718 &  & 36.225 & 7.195 & 12.813 & 0.31 \\
    8 & 1081.2 &  & 41.389 & 77.451 & 56.609 &  & $-$4.230 & $-$6.492 & $-$4.746 & 1.00 \\
    9 & 1174.2 &  & 26.523 & 41.623 & 22.759 &  & 4.429 & $-$7.444 & $-$5.942 & 0.68 \\
    10 & 1210.0 &  & 9.406 & 3.321 & $-$1.897 &  & 5.532 & 6.305 & 10.610 & $-$0.34 \\
    11 & 1216.9\smash{$^{\rm b}$} &  & 7.128 & 6.645 & 3.949 &  & 0.727 & $-$0.270 & $-$0.280 & 0.57 \\
    12 & 1238.0 &  & 13.718 & 13.650 & 12.534 &  & 2.675 & 5.578 & 5.592 & 0.92 \\
    13 & 1345.4 &  & 15.428 & 4.023 & $-$2.633 &  & 4.777 & $-$1.533 & $-$3.001 & $-$0.33 \\
    14 & 1446.7 &  & 10.422 & 20.685 & 13.452 &  & $-$1.820 & $-$0.449 & $-$0.319 & 0.92 \\
    15 & 1506.4 &  & 38.124 & 60.533 & 44.963 &  & $-$4.836 & $-$1.675 & $-$1.329 & 0.94 \\
    16 & 1522.9 &  & 13.163 & 11.464 & 12.133 &  & $-$1.618 & $-$1.344 & $-$1.440 & 0.99 \\
    17 & 1538.2 &  & 13.470 & 22.412 & 16.527 &  & $-$1.686 & $-$3.391 & $-$2.629 & 0.95 \\
    18 & 1606.7 &  & 27.024 & 10.419 & 15.438 &  & $-$5.744 & $-$1.137 & $-$1.831 & 0.92 \\
    19 & 3142.4 &  & 24.493 & 74.329 & 42.549 &  & $-$1.028 & $-$2.129 & $-$1.222 & 1.00 \\
    20 & 3205.7 &  & 29.178 & 99.811 & 53.933 &  & 2.572 & 3.143 & 1.699 & 1.00 \\
    21 & 3225.3\smash{$^{\rm a}$} &  & 25.199 & 142.897 & 59.988 &  & $-$4.484 & $-$3.932 & $-$1.651 & 1.00 \\
    22 & 3221.8\smash{$^{\rm a}$} &  & 14.148 & 31.706 & 20.754 &  & 13.677 & 21.756 & 14.533 & 0.98 \\
    23 & 3243.3 &  & 41.246 & 204.485 & 91.816 &  & $-$9.278 & $-$18.982 & $-$8.525 & 1.00 \\
    24 & 3301.8 &  & 37.907 & 147.270 & 74.679 &  & 4.429 & 10.952 & 5.556 & 1.00 \\
    \cmidrule{8-10}
    \multicolumn{3}{l}{RMS dev.\ from VG} &  &  &  &  & 11.30 & \multicolumn{1}{c}{---} & 5.61 &  \\
    \multicolumn{3}{l}{$N_{\rm sign}$ LG/VG} &  &  &  &  & \multicolumn{3}{c}{4} &  \\
    \cmidrule{1-11}
    \multicolumn{11}{c}{CISD/aug-cc-pVDZ} \\
    \cmidrule{1-11}
    1 & 216.2 &  & 8.546 & 8.615 & 6.965 &  & $-$3.332 & $-$6.167 & $-$6.142 & 0.81 \\
    2 & 379.7 &  & 40.387 & 38.611 & 2.618 &  & 13.836 & 10.931 & 11.179 & 0.07 \\
    3 & 422.2 &  & 47.486 & 47.227 & 33.420 &  & 6.286 & $-$0.606 & $-$0.608 & 0.71 \\
    4 & 813.1 &  & 34.012 & 18.854 & 19.959 &  & $-$7.652 & 5.888 & 7.908 & 0.79 \\
    5 & 896.7 &  & 205.808 & 226.110 & 212.128 &  & $-$1.955 & 4.250 & 4.055 & 0.98 \\
    6 & 934.0 &  & 13.987 & 47.162 & 19.065 &  & $-$13.407 & $-$22.701 & $-$12.362 & 0.74 \\
    7 & 1018.1 &  & 58.134 & 61.555 & 50.285 &  & 16.975 & $-$1.783 & $-$1.733 & 0.84 \\
    8 & 1075.7 &  & 37.267 & 1.888 & 8.321 &  & $-$10.760 & $-$2.109 & $-$9.369 & 0.99 \\
    9 & 1162.2 &  & 28.951 & 26.609 & 26.390 &  & 9.577 & 11.360 & 11.849 & 0.95 \\
    10 & 1210.2\smash{$^{\rm ab}$} &  & 13.955 & 76.479 & 26.768 &  & 7.000 & 24.205 & 10.339 & 0.82 \\
    11 & 1207.1\smash{$^{\rm ab}$} &  & 3.422 & 15.489 & 1.085 &  & $-$2.471 & $-$5.616 & $-$2.640 & 0.15 \\
    12 & 1230.9 &  & 6.656 & 23.826 & 11.355 &  & $-$1.067 & $-$1.573 & $-$0.831 & 0.90 \\
    13 & 1338.5 &  & 15.396 & 11.685 & $-$1.286 &  & 5.579 & $-$4.854 & $-$5.572 & $-$0.10 \\
    14 & 1441.8 &  & 8.239 & 78.284 & 4.372 &  & 0.907 & $-$6.482 & $-$2.103 & 0.17 \\
    15 & 1500.2 &  & 39.499 & 34.636 & 34.387 &  & $-$3.154 & $-$3.974 & $-$4.244 & 0.93 \\
    16 & 1519.7 &  & 12.054 & 10.303 & $-$5.784 &  & 0.509 & $-$1.147 & $-$1.240 & $-$0.52 \\
    17 & 1536.1 &  & 14.406 & 8.991 & 1.301 &  & $-$5.207 & $-$0.460 & $-$0.583 & 0.11 \\
    18 & 1594.5 &  & 25.446 & 25.840 & 22.535 &  & $-$7.192 & $-$6.182 & $-$6.135 & 0.88 \\
    19 & 3127.8 &  & 23.127 & 57.284 & 36.365 &  & $-$1.204 & $-$2.179 & $-$1.384 & 1.00 \\
    20 & 3206.9\smash{$^{\rm a}$} &  & 25.980 & 59.663 & 39.241 &  & $-$8.401 & $-$9.393 & $-$6.198 & 1.00 \\
    21 & 3204.8\smash{$^{\rm a}$} &  & 28.003 & 106.312 & 54.550 &  & 3.407 & 5.771 & 2.962 & 1.00 \\
    22 & 3215.6 &  & 10.475 & 36.884 & 19.099 &  & 19.397 & 20.434 & 10.889 & 0.97 \\
    23 & 3234.9 &  & 48.780 & 152.349 & 85.978 &  & $-$15.231 & $-$15.075 & $-$8.530 & 1.00 \\
    24 & 3303.7 &  & 28.418 & 111.509 & 56.248 &  & 5.179 & 8.335 & 4.208 & 1.00 \\
    \cmidrule{8-10}
    \multicolumn{3}{l}{RMS dev.\ from VG} &  &  &  &  & 7.41 & \multicolumn{1}{c}{---} & 4.81 &  \\
    \multicolumn{3}{l}{$N_{\rm sign}$ LG/VG} &  &  &  &  & \multicolumn{3}{c}{7} &  \\
    \cmidrule{1-11}
    \multicolumn{11}{c}{CISD/cc-pVTZ} \\
    \cmidrule{1-11}
    1 & 221.4 &  & 9.307 & 8.540 & 8.446 &  & $-$3.747 & $-$6.155 & $-$6.425 & 0.95 \\
    2 & 383.3 &  & 34.650 & 15.204 & 17.360 &  & 13.050 & 9.594 & 14.483 & 0.76 \\
    3 & 426.3 &  & 50.608 & 23.078 & 32.756 &  & 5.690 & 3.028 & 4.484 & 0.96 \\
    4 & 835.5 &  & 30.344 & 25.028 & 25.467 &  & $-$6.952 & $-$1.248 & $-$1.374 & 0.92 \\
    5 & 920.9 &  & 176.679 & 226.426 & 197.869 &  & $-$5.189 & 2.943 & 2.600 & 0.99 \\
    6 & 953.6 &  & 13.290 & 27.193 & 16.592 &  & $-$20.514 & $-$33.185 & $-$23.200 & 0.87 \\
    7 & 1045.2 &  & 62.003 & 58.238 & 56.196 &  & 24.683 & 13.705 & 14.141 & 0.94 \\
    8 & 1099.3 &  & 40.988 & 26.215 & 32.756 &  & $-$10.242 & $-$7.802 & $-$9.756 & 1.00 \\
    9 & 1191.9 &  & 25.799 & 24.428 & 25.090 &  & 7.944 & 7.458 & 7.664 & 1.00 \\
    10 & 1228.0 &  & 14.838 & 12.789 & 12.791 &  & 5.904 & 10.586 & 11.402 & 0.93 \\
    11 & 1236.6 &  & 2.245 & 16.342 & 5.524 &  & 0.968 & 2.730 & 1.012 & 0.91 \\
    12 & 1257.2 &  & 9.903 & 19.600 & 13.782 &  & 2.817 & 3.397 & 2.414 & 0.99 \\
    13 & 1363.4 &  & 17.915 & 5.788 & 8.714 &  & 7.085 & 2.650 & 4.662 & 0.86 \\
    14 & 1468.6 &  & 9.781 & 20.322 & 7.844 &  & $-$0.836 & $-$2.147 & $-$1.490 & 0.56 \\
    15 & 1524.2 &  & 46.539 & 46.021 & 45.610 &  & $-$7.578 & $-$6.321 & $-$6.357 & 0.99 \\
    16 & 1550.4 &  & 13.179 & 3.591 & 4.971 &  & $-$1.566 & $-$1.500 & $-$2.873 & 0.72 \\
    17 & 1565.1 &  & 13.973 & 7.582 & 8.716 &  & $-$2.105 & $-$2.033 & $-$2.759 & 0.85 \\
    18 & 1620.0 &  & 21.669 & 20.886 & 20.215 &  & $-$6.633 & $-$6.716 & $-$6.841 & 0.95 \\
    19 & 3144.4 &  & 20.077 & 42.114 & 29.051 &  & $-$0.870 & $-$1.458 & $-$1.007 & 1.00 \\
    20 & 3215.8 &  & 22.585 & 45.611 & 32.093 &  & 1.730 & 2.780 & 1.957 & 1.00 \\
    21 & 3216.6 &  & 30.083 & 84.061 & 50.273 &  & $-$2.066 & $-$2.392 & $-$1.431 & 1.00 \\
    22 & 3224.4 &  & 7.071 & 18.637 & 11.175 &  & 12.557 & 15.670 & 9.652 & 0.97 \\
    23 & 3243.5 &  & 46.953 & 114.680 & 73.353 &  & $-$11.014 & $-$13.994 & $-$8.955 & 1.00 \\
    24 & 3307.1 &  & 29.882 & 78.168 & 48.316 &  & 4.441 & 7.258 & 4.487 & 1.00 \\
    \cmidrule{8-10}
    \multicolumn{3}{l}{RMS dev.\ from VG} &  &  &  &  & 4.51 & \multicolumn{1}{c}{---} & 2.98 &  \\
    \multicolumn{3}{l}{$N_{\rm sign}$ LG/VG} &  &  &  &  & \multicolumn{3}{c}{1} &  \\
    \bottomrule
    \end{tabular*}
    \par\vspace{2pt}\noindent\parbox{\textwidth}{\raggedright
    $^{\rm a}$Frequency order differs from the reference level (CISD/cc-pVTZ); the row follows the matched normal mode.
    $^{\rm b}$Normal-mode overlap with the reference mode $<0.80$.
    RMS: root-mean-square difference of the rotatory strengths over all modes. $N_{\rm sign}$: number of modes whose LG and VG rotatory strengths differ in sign, counted for modes with $\max(|R_1|,|R_2|)\geq1\times10^{-44}$ esu$^2$ cm$^2$.}

\end{table}

\clearpage
\section{($S$)-3-Chloro-1-butyne}
\label{si:cby}

\begin{figure}[htbp]
\centering
\includegraphics[width=\textwidth]{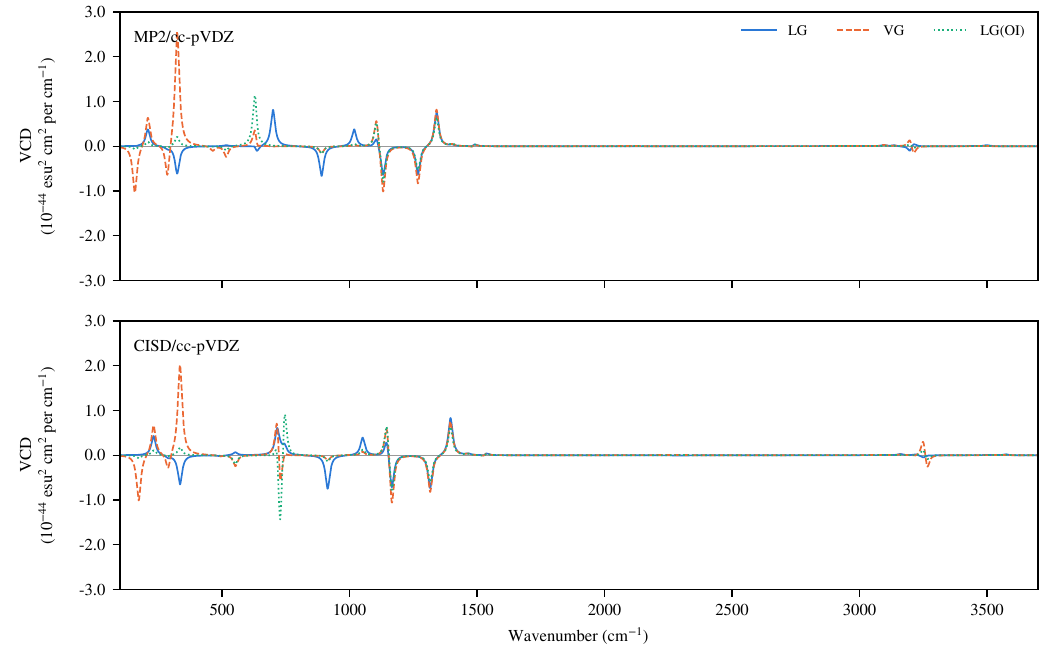}
\caption{MP2 (top) and CISD (bottom) VCD spectra of ($S$)-3-chloro-1-butyne computed with the cc-pVDZ basis set,
showing the LG (blue, solid), VG (orange, dashed), and LG(OI) (green,
dotted) rotatory strengths. Each band is broadened with an area-normalized
Lorentzian function with a half-width at half-maximum of 10~cm$^{-1}$.}
\label{fig:si_cby_pvdz}
\end{figure}

\begin{figure}[htbp]
\centering
\includegraphics[width=\textwidth]{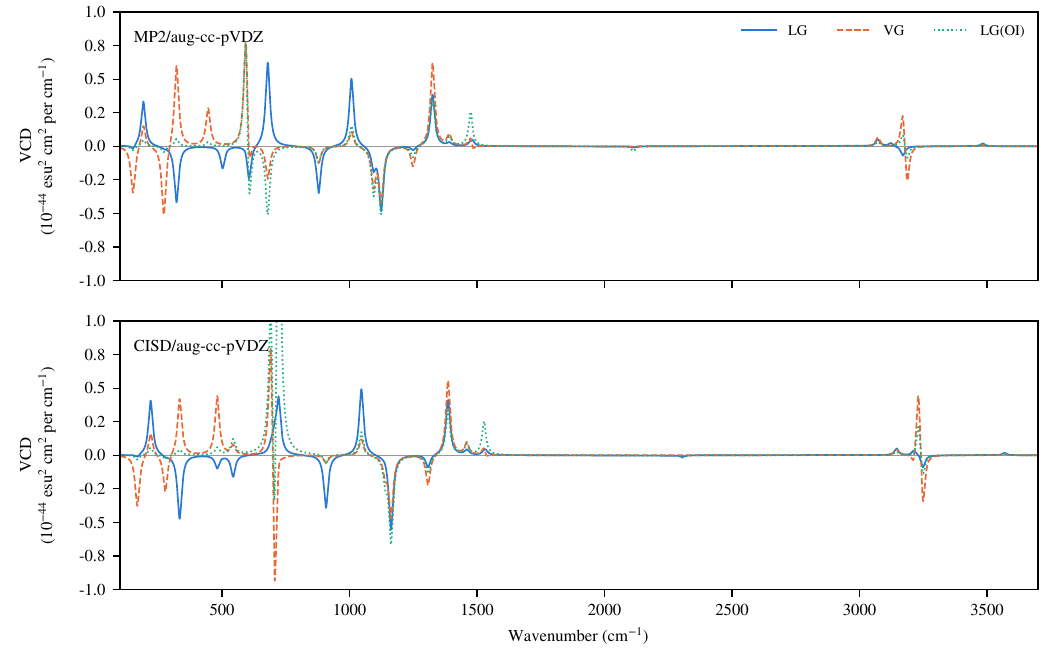}
\caption{MP2 (top) and CISD (bottom) VCD spectra of ($S$)-3-chloro-1-butyne computed with the aug-cc-pVDZ basis set,
showing the LG (blue, solid), VG (orange, dashed), and LG(OI) (green,
dotted) rotatory strengths. Each band is broadened with an area-normalized
Lorentzian function with a half-width at half-maximum of 10~cm$^{-1}$.
The CISD LG(OI) band near 722~cm$^{-1}$, whose maximum is approximately 2.4
in the units of the vertical axis, extends beyond the plotted range.}
\label{fig:si_cby_apvdz}
\end{figure}

\begin{table}[htbp]
\caption{MP2 VCD harmonic vibrational frequencies (cm$^{-1}$), LG, VG, and
mixed LG/VG dipole strengths (10$^{-40}$ esu$^{2}$ cm$^{2}$), LG, VG, and
LG(OI) rotatory strengths (10$^{-44}$ esu$^{2}$ cm$^{2}$), and the cosine of
the angle between the LG and VG APT vectors, $\cos\theta$, with respect to
basis set for each of the 24 vibrational modes of ($S$)-3-chloro-1-butyne. LG rotatory
strengths were computed with the molecular center of mass as the gauge
origin. Modes are numbered by ascending frequency at the CISD/cc-pVTZ
level.}
\label{tab:si_cby_mp2}
\scriptsize
    \renewcommand{\arraystretch}{0.6}
    \begin{tabular*}{\textwidth}{@{\extracolsep{\fill}}lrcrrrcrrrr}
    \toprule
    & & & \multicolumn{3}{c}{Dipole Strength} & \hspace{1pt} & \multicolumn{3}{c}{Rotatory Strength} & \\
    \cmidrule{4-6} \cmidrule{8-10}
    Mode & \multicolumn{1}{c}{$\nu$} &  & \multicolumn{1}{c}{LG} & \multicolumn{1}{c}{VG} & \multicolumn{1}{c}{Mixed} &  & \multicolumn{1}{c}{LG} & \multicolumn{1}{c}{VG} & \multicolumn{1}{c}{LG(OI)} & \multicolumn{1}{c}{$\cos\theta$} \\ \midrule
    \multicolumn{11}{c}{MP2/cc-pVDZ} \\
    \cmidrule{1-11}
    1 & 158.2 &  & 30.850 & 6642.584 & $-$285.186 &  & $-$0.343 & $-$33.421 & $-$2.278 & $-$0.63 \\
    2 & 209.2 &  & 28.126 & 1251.608 & 18.018 &  & 12.030 & 20.989 & 3.146 & 0.10 \\
    3 & 285.8 &  & 2.600 & 556.053 & 4.302 &  & $-$1.271 & $-$25.338 & $-$1.733 & 0.11 \\
    4 & 324.2 &  & 18.429 & 2843.290 & $-$188.279 &  & $-$19.234 & 82.212 & 6.619 & $-$0.82 \\
    5 & 462.8 &  & 15.307 & 1523.727 & $-$118.257 &  & 0.074 & $-$3.434 & $-$0.344 & $-$0.77 \\
    6 & 516.4 &  & 38.950 & 212.184 & $-$63.849 &  & 0.830 & $-$7.591 & $-$3.252 & $-$0.70 \\
    7 & 700.0\smash{$^{\rm a}$} &  & 219.988 & 292.534 & $-$233.239 &  & 25.755 & $-$0.309 & $-$0.268 & $-$0.92 \\
    8 & 629.8\smash{$^{\rm a}$} &  & 238.718 & 38.080 & 72.874 &  & 2.510 & 16.988 & 42.534 & 0.76 \\
    9 & 636.0\smash{$^{\rm a}$} &  & 136.130 & 110.801 & 116.271 &  & $-$5.501 & $-$8.776 & $-$9.727 & 0.95 \\
    10 & 890.3 &  & 67.968 & 63.506 & 53.067 &  & $-$20.974 & $-$5.178 & $-$5.357 & 0.81 \\
    11 & 1018.1 &  & 51.366 & 21.925 & 22.698 &  & 12.087 & 1.064 & 1.628 & 0.68 \\
    12 & 1105.8 &  & 72.045 & 88.450 & 73.201 &  & 7.416 & 22.043 & 19.894 & 0.92 \\
    13 & 1131.8 &  & 26.064 & 40.902 & 30.677 &  & $-$21.092 & $-$34.503 & $-$27.542 & 0.94 \\
    14 & 1268.4 &  & 95.117 & 234.043 & 147.046 &  & $-$20.026 & $-$26.477 & $-$16.879 & 0.99 \\
    15 & 1341.2 &  & 37.918 & 61.489 & 47.823 &  & 24.631 & 26.184 & 20.562 & 0.99 \\
    16 & 1402.8 &  & 11.461 & 4.922 & 7.291 &  & 0.827 & 1.096 & 1.672 & 0.97 \\
    17 & 1482.4 &  & 16.581 & 16.769 & 16.649 &  & $-$1.468 & $-$1.477 & $-$1.469 & 1.00 \\
    18 & 1489.0 &  & 4.520 & 4.088 & 4.271 &  & 2.108 & 2.000 & 2.103 & 0.99 \\
    19 & 2132.6 &  & 9.321 & 6.288 & 7.570 &  & $-$0.285 & 0.182 & 0.222 & 0.99 \\
    20 & 3092.6 &  & 13.245 & 72.175 & 30.757 &  & 0.904 & 1.014 & 0.434 & 0.99 \\
    21 & 3134.5 &  & 6.241 & 49.273 & 17.113 &  & 0.784 & 0.505 & 0.180 & 0.98 \\
    22 & 3197.7 &  & 11.591 & 134.476 & 39.363 &  & $-$3.973 & 6.179 & 1.814 & 1.00 \\
    23 & 3211.9 &  & 10.092 & 134.897 & 36.803 &  & 2.610 & $-$6.550 & $-$1.792 & 1.00 \\
    24 & 3496.6 &  & 73.093 & 34.640 & 50.287 &  & 0.849 & 0.302 & 0.439 & 1.00 \\
    \cmidrule{8-10}
    \multicolumn{3}{l}{RMS dev.\ from VG} &  &  &  &  & 24.17 & \multicolumn{1}{c}{---} & 18.78 &  \\
    \multicolumn{3}{l}{$N_{\rm sign}$ LG/VG} &  &  &  &  & \multicolumn{3}{c}{6} &  \\
    \cmidrule{1-11}
    \multicolumn{11}{c}{MP2/aug-cc-pVDZ} \\
    \cmidrule{1-11}
    1 & 150.8 &  & 39.464 & 5073.155 & $-$325.062 &  & $-$1.054 & $-$11.131 & $-$0.982 & $-$0.73 \\
    2 & 191.5 &  & 31.055 & 410.295 & 10.932 &  & 10.635 & 5.485 & 1.509 & 0.10 \\
    3 & 271.8 &  & 2.258 & 303.808 & 3.031 &  & $-$0.323 & $-$16.780 & $-$1.447 & 0.12 \\
    4 & 321.5 &  & 17.674 & 2181.835 & $-$175.122 &  & $-$13.145 & 19.362 & 1.743 & $-$0.89 \\
    5 & 445.4 &  & 17.408 & 1428.192 & $-$116.806 &  & $-$0.108 & 8.787 & 0.970 & $-$0.74 \\
    6 & 502.3 &  & 25.505 & 58.110 & $-$32.462 &  & $-$5.157 & 0.035 & 0.023 & $-$0.84 \\
    7 & 679.5\smash{$^{\rm a}$} &  & 277.740 & 66.220 & $-$121.998 &  & 19.734 & $-$7.801 & $-$15.977 & $-$0.90 \\
    8 & 605.5\smash{$^{\rm b}$} &  & 258.245 & 80.596 & 136.146 &  & $-$8.243 & $-$12.943 & $-$23.168 & 0.94 \\
    9 & 593.8\smash{$^{\rm ab}$} &  & 214.268 & 170.936 & 156.307 &  & 1.118 & 29.691 & 33.242 & 0.82 \\
    10 & 879.1 &  & 73.790 & 91.254 & 69.388 &  & $-$11.056 & $-$3.952 & $-$3.554 & 0.85 \\
    11 & 1007.6 &  & 49.896 & 25.062 & 4.070 &  & 15.990 & 3.546 & 5.003 & 0.12 \\
    12 & 1094.4 &  & 78.280 & 56.715 & 59.027 &  & $-$4.713 & $-$8.771 & $-$10.304 & 0.89 \\
    13 & 1124.2 &  & 20.790 & 14.430 & 10.894 &  & $-$14.823 & $-$12.385 & $-$14.866 & 0.63 \\
    14 & 1248.8 &  & 91.411 & 321.495 & 169.447 &  & $-$1.064 & $-$4.966 & $-$2.648 & 0.99 \\
    15 & 1325.7 &  & 33.520 & 102.821 & 55.847 &  & 12.092 & 19.575 & 11.177 & 0.95 \\
    16 & 1391.0 &  & 16.143 & 28.233 & $-$19.521 &  & 0.721 & 2.569 & 1.943 & $-$0.91 \\
    17 & 1476.2 &  & 17.160 & 2.707 & $-$0.755 &  & 0.690 & 3.955 & 9.957 & $-$0.11 \\
    18 & 1481.4 &  & 6.150 & 8.946 & $-$5.314 &  & 0.992 & $-$3.062 & $-$2.539 & $-$0.72 \\
    19 & 2113.2 &  & 4.096 & 0.043 & 0.407 &  & $-$0.340 & $-$0.118 & $-$1.161 & 0.98 \\
    20 & 3071.6 &  & 13.378 & 50.300 & 25.862 &  & 1.626 & 2.018 & 1.041 & 1.00 \\
    21 & 3122.1 &  & 2.326 & 24.462 & 6.732 &  & 0.813 & $-$0.137 & $-$0.042 & 0.89 \\
    22 & 3170.1 &  & 12.020 & 92.685 & 33.279 &  & $-$2.216 & 9.816 & 3.535 & 1.00 \\
    23 & 3186.5 &  & 10.947 & 94.805 & 32.042 &  & $-$0.031 & $-$10.611 & $-$3.606 & 0.99 \\
    24 & 3483.2 &  & 69.891 & 36.553 & 50.496 &  & 0.660 & 0.328 & 0.453 & 1.00 \\
    \cmidrule{8-10}
    \multicolumn{3}{l}{RMS dev.\ from VG} &  &  &  &  & 12.49 & \multicolumn{1}{c}{---} & 6.84 &  \\
    \multicolumn{3}{l}{$N_{\rm sign}$ LG/VG} &  &  &  &  & \multicolumn{3}{c}{6} &  \\
    \cmidrule{1-11}
    \multicolumn{11}{c}{MP2/cc-pVTZ} \\
    \cmidrule{1-11}
    1 & 159.2 &  & 33.775 & 5827.840 & $-$325.998 &  & $-$0.668 & $-$16.875 & $-$1.285 & $-$0.73 \\
    2 & 208.9 &  & 29.119 & 708.523 & 16.412 &  & 12.402 & 15.258 & 3.093 & 0.11 \\
    3 & 276.4 &  & 2.948 & 443.798 & 4.226 &  & $-$1.033 & $-$11.102 & $-$0.905 & 0.12 \\
    4 & 320.4 &  & 18.279 & 2711.431 & $-$179.652 &  & $-$14.535 & $-$1.694 & $-$0.139 & $-$0.81 \\
    5 & 467.1 &  & 17.486 & 1769.351 & $-$74.849 &  & $-$3.580 & 21.124 & 2.100 & $-$0.43 \\
    6 & 522.6 &  & 42.470 & 39.759 & $-$37.326 &  & $-$4.682 & $-$1.229 & $-$1.270 & $-$0.91 \\
    7 & 699.0\smash{$^{\rm a}$} &  & 216.998 & 188.170 & $-$193.417 &  & 25.263 & $-$13.613 & $-$14.619 & $-$0.96 \\
    8 & 633.8\smash{$^{\rm a}$} &  & 270.774 & 121.142 & 175.072 &  & 1.140 & 21.447 & 32.064 & 0.97 \\
    9 & 643.6\smash{$^{\rm a}$} &  & 161.540 & 206.795 & 181.393 &  & $-$5.990 & $-$6.940 & $-$6.134 & 0.99 \\
    10 & 887.2 &  & 67.175 & 36.396 & 33.455 &  & $-$16.744 & 2.172 & 2.951 & 0.68 \\
    11 & 1018.7 &  & 52.431 & 21.928 & 14.476 &  & 16.278 & 2.671 & 4.130 & 0.43 \\
    12 & 1105.4 &  & 57.531 & 42.722 & 47.106 &  & 5.690 & 6.947 & 8.061 & 0.95 \\
    13 & 1130.0 &  & 35.701 & 20.171 & 22.969 &  & $-$23.132 & $-$26.012 & $-$34.605 & 0.86 \\
    14 & 1268.1 &  & 82.418 & 232.554 & 136.215 &  & $-$11.306 & $-$14.812 & $-$8.818 & 0.98 \\
    15 & 1346.6 &  & 27.862 & 48.175 & 36.251 &  & 19.315 & 25.092 & 19.082 & 0.99 \\
    16 & 1409.5 &  & 15.981 & 1.826 & $-$2.621 &  & 0.546 & 1.188 & 3.515 & $-$0.49 \\
    17 & 1500.6 &  & 18.403 & 4.878 & 9.180 &  & $-$0.502 & 0.235 & 0.457 & 0.97 \\
    18 & 1506.7 &  & 4.959 & 0.530 & $-$0.358 &  & 1.494 & 0.289 & 0.884 & $-$0.22 \\
    19 & 2138.1 &  & 4.548 & 0.758 & 1.850 &  & $-$0.247 & $-$0.110 & $-$0.270 & 1.00 \\
    20 & 3084.4 &  & 11.460 & 39.955 & 21.354 &  & 1.354 & 1.791 & 0.959 & 1.00 \\
    21 & 3130.8 &  & 2.360 & 15.240 & 5.700 &  & 0.832 & 0.374 & 0.147 & 0.95 \\
    22 & 3179.2 &  & 10.366 & 65.252 & 25.953 &  & $-$1.588 & 5.516 & 2.198 & 1.00 \\
    23 & 3193.4 &  & 9.319 & 65.059 & 24.549 &  & $-$0.195 & $-$6.733 & $-$2.548 & 1.00 \\
    24 & 3497.2 &  & 72.986 & 48.234 & 59.311 &  & 0.739 & 0.431 & 0.530 & 1.00 \\
    \cmidrule{8-10}
    \multicolumn{3}{l}{RMS dev.\ from VG} &  &  &  &  & 12.55 & \multicolumn{1}{c}{---} & 6.95 &  \\
    \multicolumn{3}{l}{$N_{\rm sign}$ LG/VG} &  &  &  &  & \multicolumn{3}{c}{4} &  \\
    \bottomrule
    \end{tabular*}
    \par\vspace{2pt}\noindent\parbox{\textwidth}{\raggedright
    $^{\rm a}$Frequency order differs from the reference level (CISD/cc-pVTZ); the row follows the matched normal mode.
    $^{\rm b}$Normal-mode overlap with the reference mode $<0.80$.
    RMS: root-mean-square difference of the rotatory strengths over all modes. $N_{\rm sign}$: number of modes whose LG and VG rotatory strengths differ in sign, counted for modes with $\max(|R_1|,|R_2|)\geq1\times10^{-44}$ esu$^2$ cm$^2$.}

\end{table}

\begin{table}[htbp]
\caption{CISD VCD harmonic vibrational frequencies (cm$^{-1}$), LG, VG, and
mixed LG/VG dipole strengths (10$^{-40}$ esu$^{2}$ cm$^{2}$), LG, VG, and
LG(OI) rotatory strengths (10$^{-44}$ esu$^{2}$ cm$^{2}$), and the cosine of
the angle between the LG and VG APT vectors, $\cos\theta$, with respect to
basis set for each of the 24 vibrational modes of ($S$)-3-chloro-1-butyne. LG rotatory
strengths were computed with the molecular center of mass as the gauge
origin. Modes are numbered by ascending frequency at the CISD/cc-pVTZ
level.}
\label{tab:si_cby_cisd}
\scriptsize
    \renewcommand{\arraystretch}{0.6}
    \begin{tabular*}{\textwidth}{@{\extracolsep{\fill}}lrcrrrcrrrr}
    \toprule
    & & & \multicolumn{3}{c}{Dipole Strength} & \hspace{1pt} & \multicolumn{3}{c}{Rotatory Strength} & \\
    \cmidrule{4-6} \cmidrule{8-10}
    Mode & \multicolumn{1}{c}{$\nu$} &  & \multicolumn{1}{c}{LG} & \multicolumn{1}{c}{VG} & \multicolumn{1}{c}{Mixed} &  & \multicolumn{1}{c}{LG} & \multicolumn{1}{c}{VG} & \multicolumn{1}{c}{LG(OI)} & \multicolumn{1}{c}{$\cos\theta$} \\ \midrule
    \multicolumn{11}{c}{CISD/cc-pVDZ} \\
    \cmidrule{1-11}
    1 & 173.3 &  & 33.918 & 6431.188 & $-$301.054 &  & $-$0.003 & $-$32.489 & $-$2.359 & $-$0.64 \\
    2 & 231.5 &  & 28.422 & 1024.670 & 15.419 &  & 13.719 & 21.691 & 3.613 & 0.09 \\
    3 & 288.5 &  & 4.287 & 704.103 & 9.506 &  & $-$1.613 & $-$11.965 & $-$0.934 & 0.17 \\
    4 & 335.2 &  & 20.163 & 2723.379 & $-$180.397 &  & $-$20.540 & 63.970 & 5.504 & $-$0.77 \\
    5 & 492.7 &  & 18.580 & 1468.265 & $-$107.611 &  & $-$0.815 & $-$0.708 & $-$0.080 & $-$0.65 \\
    6 & 552.2 &  & 78.418 & 127.984 & $-$75.881 &  & 2.092 & $-$7.868 & $-$6.159 & $-$0.76 \\
    7 & 716.3 &  & 151.753 & 363.300 & $-$134.980 &  & 17.616 & 34.257 & 22.140 & $-$0.57 \\
    8 & 727.9 &  & 176.180 & 43.081 & 16.196 &  & 2.192 & $-$32.072 & $-$64.857 & 0.19 \\
    9 & 745.2 &  & 239.202 & 2.685 & 21.339 &  & 5.806 & 4.411 & 41.633 & 0.84 \\
    10 & 914.3 &  & 84.801 & 55.778 & 52.085 &  & $-$23.626 & $-$3.613 & $-$4.456 & 0.76 \\
    11 & 1051.5 &  & 52.210 & 20.337 & 21.388 &  & 12.553 & 2.453 & 3.930 & 0.66 \\
    12 & 1145.8 &  & 65.099 & 65.757 & 58.592 &  & 13.572 & 26.127 & 25.996 & 0.90 \\
    13 & 1165.8 &  & 31.518 & 49.179 & 38.338 &  & $-$25.099 & $-$38.259 & $-$30.629 & 0.97 \\
    14 & 1316.2 &  & 111.097 & 208.657 & 149.851 &  & $-$23.407 & $-$26.092 & $-$19.039 & 0.98 \\
    15 & 1396.1 &  & 37.946 & 53.795 & 44.433 &  & 26.505 & 24.108 & 20.247 & 0.98 \\
    16 & 1464.0 &  & 8.500 & 8.393 & 8.235 &  & 1.040 & 0.880 & 0.886 & 0.97 \\
    17 & 1528.7 &  & 15.069 & 18.511 & 16.691 &  & $-$1.583 & $-$1.618 & $-$1.460 & 1.00 \\
    18 & 1535.0 &  & 3.749 & 5.885 & 4.682 &  & 2.177 & 2.220 & 1.772 & 1.00 \\
    19 & 2315.3 &  & 4.724 & 4.478 & 4.546 &  & $-$0.383 & 0.247 & 0.254 & 0.99 \\
    20 & 3158.1 &  & 15.877 & 62.431 & 31.301 &  & 0.994 & 0.537 & 0.271 & 0.99 \\
    21 & 3219.9 &  & 4.228 & 29.454 & 10.681 &  & 0.789 & $-$0.867 & $-$0.328 & 0.96 \\
    22 & 3250.0 &  & 16.221 & 118.934 & 43.822 &  & $-$1.631 & 12.792 & 4.724 & 1.00 \\
    23 & 3265.4 &  & 14.932 & 124.609 & 43.039 &  & 0.120 & $-$11.574 & $-$4.007 & 1.00 \\
    24 & 3573.7 &  & 74.904 & 30.434 & 47.731 &  & 0.685 & 0.242 & 0.380 & 1.00 \\
    \cmidrule{8-10}
    \multicolumn{3}{l}{RMS dev.\ from VG} &  &  &  &  & 21.52 & \multicolumn{1}{c}{---} & 17.83 &  \\
    \multicolumn{3}{l}{$N_{\rm sign}$ LG/VG} &  &  &  &  & \multicolumn{3}{c}{5} &  \\
    \cmidrule{1-11}
    \multicolumn{11}{c}{CISD/aug-cc-pVDZ} \\
    \cmidrule{1-11}
    1 & 167.8 &  & 42.045 & 4855.081 & $-$330.993 &  & $-$0.768 & $-$12.065 & $-$1.123 & $-$0.73 \\
    2 & 220.1 &  & 32.405 & 327.849 & 6.923 &  & 12.952 & 5.614 & 1.765 & 0.07 \\
    3 & 277.6 &  & 4.446 & 358.927 & 9.570 &  & $-$0.474 & $-$8.968 & $-$0.998 & 0.24 \\
    4 & 333.8 &  & 19.678 & 2147.900 & $-$175.103 &  & $-$14.938 & 13.368 & 1.279 & $-$0.85 \\
    5 & 481.0 &  & 19.407 & 1369.177 & $-$72.061 &  & $-$2.981 & 13.777 & 1.640 & $-$0.44 \\
    6 & 543.6 &  & 69.796 & 26.479 & $-$41.640 &  & $-$5.023 & 2.168 & 3.519 & $-$0.97 \\
    7 & 691.9\smash{$^{\rm b}$} &  & 162.240 & 141.337 & 51.957 &  & 1.377 & 37.919 & 40.626 & 0.34 \\
    8 & 706.0\smash{$^{\rm b}$} &  & 123.019 & 96.111 & 15.547 &  & 3.549 & $-$42.884 & $-$48.517 & 0.14 \\
    9 & 721.7 &  & 389.032 & 1.228 & 3.725 &  & 12.691 & 4.804 & 85.510 & 0.17 \\
    10 & 907.7 &  & 87.970 & 73.111 & 64.361 &  & $-$12.419 & $-$1.837 & $-$2.015 & 0.80 \\
    11 & 1046.6 &  & 49.564 & 24.472 & 1.880 &  & 15.624 & 3.896 & 5.544 & 0.05 \\
    12 & 1140.2 &  & 72.002 & 42.828 & 47.894 &  & 0.466 & $-$3.833 & $-$4.969 & 0.86 \\
    13 & 1162.7 &  & 24.549 & 12.167 & 10.222 &  & $-$17.685 & $-$14.175 & $-$20.135 & 0.59 \\
    14 & 1308.0 &  & 108.043 & 288.093 & 174.230 &  & $-$2.945 & $-$7.191 & $-$4.404 & 0.99 \\
    15 & 1386.2 &  & 33.419 & 92.389 & 51.536 &  & 12.976 & 17.585 & 10.576 & 0.93 \\
    16 & 1459.6 &  & 12.239 & 17.743 & $-$13.278 &  & 1.160 & 2.835 & 2.355 & $-$0.90 \\
    17 & 1528.6 &  & 15.919 & 1.794 & 0.380 &  & 0.636 & 3.104 & 9.247 & 0.07 \\
    18 & 1533.6 &  & 4.685 & 6.831 & $-$4.120 &  & 0.833 & $-$2.193 & $-$1.816 & $-$0.73 \\
    19 & 2304.1 &  & 1.431 & 0.002 & 0.047 &  & $-$0.467 & $-$0.019 & $-$0.577 & 0.97 \\
    20 & 3145.3 &  & 15.403 & 43.346 & 25.752 &  & 1.608 & 1.304 & 0.777 & 1.00 \\
    21 & 3213.1 &  & 0.873 & 9.329 & 1.986 &  & 1.174 & $-$3.911 & $-$1.196 & 0.70 \\
    22 & 3231.3 &  & 17.791 & 84.034 & 38.527 &  & 0.524 & 18.412 & 8.472 & 1.00 \\
    23 & 3248.5 &  & 15.779 & 88.213 & 37.098 &  & $-$3.057 & $-$14.874 & $-$6.291 & 0.99 \\
    24 & 3568.3 &  & 73.603 & 31.041 & 47.774 &  & 0.536 & 0.224 & 0.345 & 1.00 \\
    \cmidrule{8-10}
    \multicolumn{3}{l}{RMS dev.\ from VG} &  &  &  &  & 15.50 & \multicolumn{1}{c}{---} & 17.51 &  \\
    \multicolumn{3}{l}{$N_{\rm sign}$ LG/VG} &  &  &  &  & \multicolumn{3}{c}{7} &  \\
    \cmidrule{1-11}
    \multicolumn{11}{c}{CISD/cc-pVTZ} \\
    \cmidrule{1-11}
    1 & 175.9 &  & 36.464 & 5593.322 & $-$333.329 &  & $-$0.332 & $-$17.149 & $-$1.385 & $-$0.74 \\
    2 & 232.8 &  & 28.921 & 505.079 & 11.801 &  & 14.394 & 11.052 & 2.645 & 0.10 \\
    3 & 282.8 &  & 5.868 & 553.338 & 13.041 &  & $-$1.344 & $-$0.167 & $-$0.017 & 0.23 \\
    4 & 334.6 &  & 19.770 & 2636.799 & $-$173.208 &  & $-$16.101 & $-$6.947 & $-$0.602 & $-$0.76 \\
    5 & 500.3 &  & 20.484 & 1727.784 & $-$51.620 &  & $-$5.621 & 25.174 & 2.741 & $-$0.27 \\
    6 & 562.3 &  & 84.946 & 9.590 & $-$26.534 &  & $-$3.925 & $-$3.125 & $-$9.300 & $-$0.93 \\
    7 & 726.7 &  & 171.168 & 260.790 & $-$183.601 &  & 19.136 & 11.349 & 9.195 & $-$0.87 \\
    8 & 747.0 &  & 218.497 & 88.102 & 134.344 &  & 4.368 & $-$15.347 & $-$24.169 & 0.97 \\
    9 & 765.5 &  & 207.646 & 64.628 & 115.748 &  & 0.687 & 0.141 & 0.253 & 1.00 \\
    10 & 917.3 &  & 83.346 & 29.626 & 29.032 &  & $-$18.531 & 4.442 & 7.450 & 0.58 \\
    11 & 1060.1 &  & 52.361 & 20.556 & 12.496 &  & 16.230 & 3.292 & 5.254 & 0.38 \\
    12 & 1155.0 &  & 43.558 & 27.638 & 31.684 &  & 11.338 & 10.423 & 13.085 & 0.91 \\
    13 & 1174.1 &  & 48.569 & 23.655 & 30.377 &  & $-$26.722 & $-$28.235 & $-$40.459 & 0.90 \\
    14 & 1333.3 &  & 98.414 & 209.040 & 141.059 &  & $-$14.422 & $-$16.215 & $-$11.126 & 0.98 \\
    15 & 1413.8 &  & 28.214 & 41.963 & 33.827 &  & 21.771 & 23.633 & 19.378 & 0.98 \\
    16 & 1483.9 &  & 12.065 & 0.665 & 0.032 &  & 0.679 & 1.291 & 5.499 & 0.01 \\
    17 & 1557.7 &  & 16.882 & 5.666 & 9.637 &  & $-$0.504 & $-$0.002 & $-$0.004 & 0.99 \\
    18 & 1563.6 &  & 4.157 & 0.288 & 0.314 &  & 1.411 & 0.555 & 2.109 & 0.29 \\
    19 & 2329.6 &  & 1.523 & 0.600 & 0.944 &  & $-$0.398 & $-$0.059 & $-$0.095 & 0.99 \\
    20 & 3160.6 &  & 13.916 & 33.047 & 21.407 &  & 1.379 & 1.345 & 0.873 & 1.00 \\
    21 & 3224.8 &  & 0.716 & 3.701 & 1.339 &  & 1.130 & $-$1.527 & $-$0.672 & 0.82 \\
    22 & 3242.1 &  & 16.498 & 56.285 & 30.417 &  & 1.569 & 11.468 & 6.209 & 1.00 \\
    23 & 3257.8 &  & 14.570 & 56.972 & 28.744 &  & $-$3.637 & $-$10.426 & $-$5.273 & 1.00 \\
    24 & 3588.7 &  & 75.477 & 44.752 & 58.112 &  & 0.605 & 0.365 & 0.473 & 1.00 \\
    \cmidrule{8-10}
    \multicolumn{3}{l}{RMS dev.\ from VG} &  &  &  &  & 10.48 & \multicolumn{1}{c}{---} & 7.28 &  \\
    \multicolumn{3}{l}{$N_{\rm sign}$ LG/VG} &  &  &  &  & \multicolumn{3}{c}{4} &  \\
    \bottomrule
    \end{tabular*}
    \par\vspace{2pt}\noindent\parbox{\textwidth}{\raggedright
    $^{\rm a}$Frequency order differs from the reference level (CISD/cc-pVTZ); the row follows the matched normal mode.
    $^{\rm b}$Normal-mode overlap with the reference mode $<0.80$.
    RMS: root-mean-square difference of the rotatory strengths over all modes. $N_{\rm sign}$: number of modes whose LG and VG rotatory strengths differ in sign, counted for modes with $\max(|R_1|,|R_2|)\geq1\times10^{-44}$ esu$^2$ cm$^2$.}

\end{table}

\clearpage
\section{Charge-sum rule for the VG-APT}
\label{si:sumrule}

\begin{table}[htbp]
\caption{Largest absolute element (in units of $e$) of the sum over all nuclei of the
VG-APT, $\sum_\lambda P^\lambda_{\alpha\beta}$, which the charge-sum rule requires to
vanish for a neutral molecule, for the full VG-APT (nuclear, reference, and correlation
contributions) and for the correlation contribution alone. The corresponding largest element
for the LG-APT is below $10^{-11}$~$e$ in every calculation. Methyloxirane results with the
cc-pVDZ and aug-cc-pVDZ basis sets were obtained for the ($R$) enantiomer; the sum is
unaffected by reflection.}
\label{tab:si_sumrule}
\small
    \begin{tabular*}{\textwidth}{@{\extracolsep{\fill}}llrrrrrr}
    \toprule
    Molecule & Method & \multicolumn{1}{c}{cc-pVDZ} & \multicolumn{1}{c}{aug-cc-pVDZ} & \multicolumn{1}{c}{cc-pVTZ} & \multicolumn{1}{c}{aug-cc-pVTZ} & \multicolumn{1}{c}{cc-pVQZ} & \multicolumn{1}{c}{aug-cc-pVQZ} \\
    \midrule
    \multicolumn{8}{c}{Full VG-APT} \\
    \cmidrule{1-8}
    H$_2$O$_2$ & MP2 & 7.58 & 4.39 & 3.01 & 1.97 & 1.13 & 0.72 \\
     & CISD & 7.56 & 4.50 & 3.04 & 2.01 & 1.14 & 0.73 \\
    methyloxirane & MP2 & 9.63 & 6.80 & 4.20 & --- & --- & --- \\
     & CISD & 9.87 & 7.01 & 4.22 & --- & --- & --- \\
    3-chloro-1-butyne & MP2 & 21.80 & 16.80 & 13.31 & --- & --- & --- \\
     & CISD & 21.85 & 16.99 & 13.34 & --- & --- & --- \\
    \midrule
    \multicolumn{8}{c}{Correlation contribution} \\
    \cmidrule{1-8}
    H$_2$O$_2$ & MP2 & 0.06 & 0.07 & 0.04 & 0.02 & 0.02 & 0.00 \\
     & CISD & 0.14 & 0.20 & 0.10 & 0.06 & 0.05 & 0.01 \\
    methyloxirane & MP2 & 0.10 & 0.09 & 0.04 & --- & --- & --- \\
     & CISD & 0.23 & 0.25 & 0.09 & --- & --- & --- \\
    3-chloro-1-butyne & MP2 & 0.13 & 0.15 & 0.05 & --- & --- & --- \\
     & CISD & 0.29 & 0.35 & 0.11 & --- & --- & --- \\
    \bottomrule
    \end{tabular*}

\end{table}